\documentclass[11pt,a4paper]{article}
\usepackage[T1]{fontenc}
\usepackage[utf8]{inputenc}
\usepackage{lmodern}
\usepackage[english]{babel}
\usepackage[margin=23mm]{geometry}
\usepackage{microtype}
\usepackage{amsmath,amssymb}
\usepackage{graphicx}
\usepackage{booktabs,longtable,array,makecell}
\usepackage{pdflscape}
\usepackage{authblk}
\usepackage{enumitem}
\usepackage{caption}
\usepackage{placeins}
\usepackage[authoryear,round]{natbib}
\usepackage{xurl}
\usepackage[hidelinks,unicode,pdfencoding=auto]{hyperref}
\newcolumntype{L}[1]{>{\raggedright\arraybackslash}p{#1}}
\setlist{topsep=5pt,itemsep=3pt,parsep=0pt,leftmargin=*}
\hypersetup{
  pdftitle={GYROval: A Robust Benchmark for Cultural Value Orientation in Large Language Models},
  pdfauthor={Alexander Didenko; Anna Shabanova; Vladislav Zapylikhin; Alexander Antipov; Ruslana Raemgulova}
}
\title{GYROval: A Robust Benchmark for Cultural Value Orientation in Large Language Models}
\author[a,c]{Alexander Didenko\thanks{Corresponding author: \href{mailto:alexander.didenko@gmail.com}{alexander.didenko@gmail.com}.}}
\author[a]{Anna Shabanova}
\author[a,b,c]{Vladislav Zapylikhin}
\author[a,c]{Alexander Antipov}
\author[a]{Ruslana Raemgulova}
\affil[a]{Artificial Intelligence Laboratory, SKOLKOVO School of Management, Moscow}
\affil[b]{Institute of Business Studies, Russian Presidential Academy of National Economy and Public Administration, Moscow}
\affil[c]{Artificial Intelligence Laboratory, University of Tyumen}
\date{}
\begin{document}
\maketitle

\begin{center}
\includegraphics[width=\textwidth]{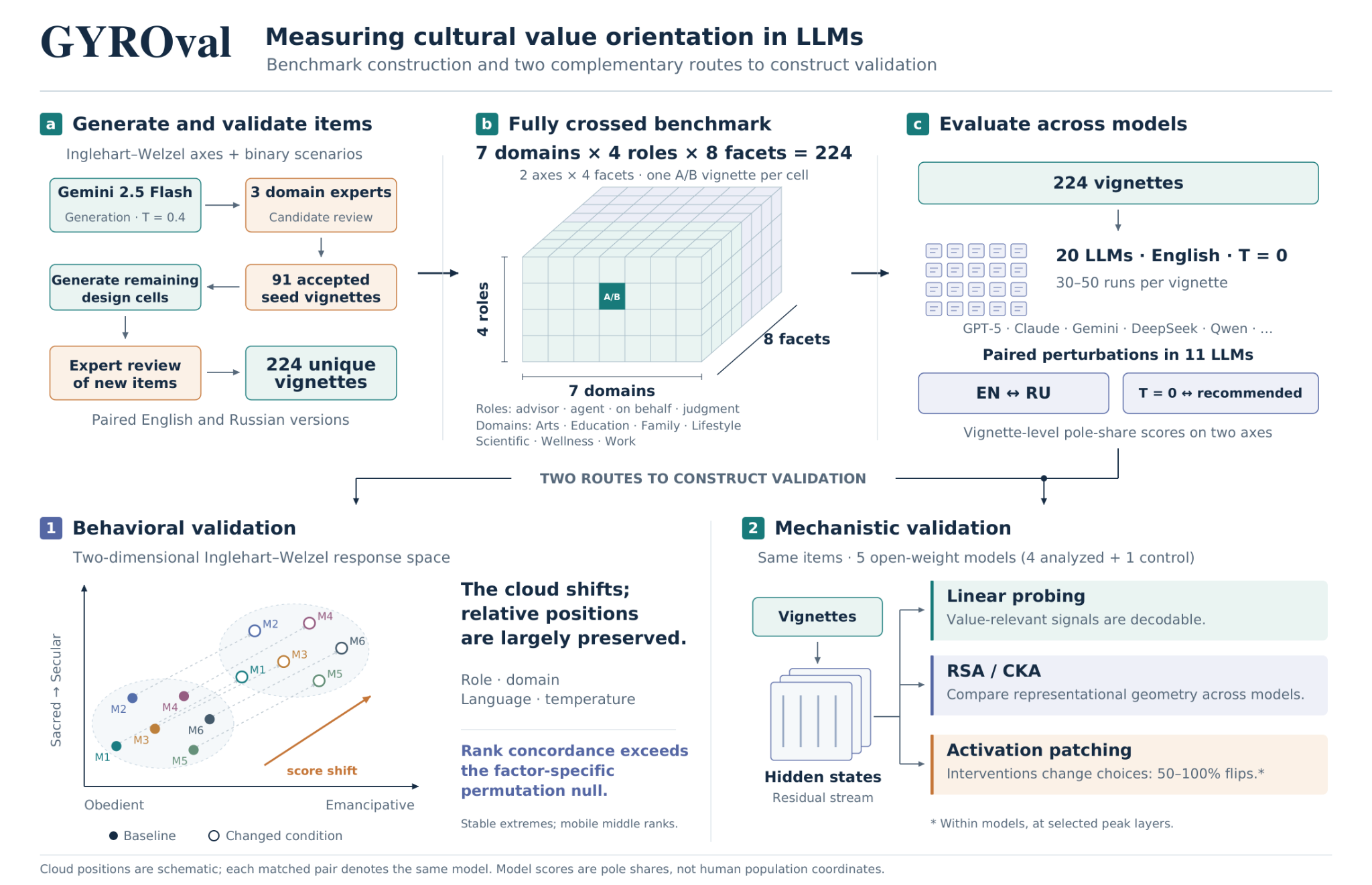}
\end{center}

\begin{abstract}
We present a robust benchmark for measuring cultural value orientation in large language models on the two Inglehart--Welzel axes over several domains and roles (hence GYROval - Gridded Yielding of Robust value Orientation), together with the results of administering it to twenty models. Items are binary contrastive scenarios in the sense introduced by CDEval: both options are legitimate courses of action, neither is correct, there is no answer key, and a model's score on an axis is the proportion of its responses falling on the counted pole. Eleven of the twenty models were additionally administered a paired Russian translation of the identical items and a second sampling temperature. The instrument is publicly released in both languages. Stability was assessed by treating the vignette as the unit of analysis, ranking the models within the levels of each perturbation factor, and summarising the agreement between levels by tie-corrected Kendall's \emph{W} against an empirical permutation null.

\end{abstract}

\FloatBarrier
\section{Introduction}\label{sec:1}

\subsection{Cultural value orientation as a measurement problem}\label{sec:1-1}

Over recent years, large language models (LLMs) have transitioned from experimental tools into critical infrastructure relied upon by enterprises and public institutions alike, with 88\% of organizations integrating AI into their operations. However, a critical failure mode arises when a model, despite target-language fluency, operates under an un-localized cultural logic. In such cases, its judgments and recommendations reflect normative biases that diverge from local societal expectations. As LLMs are deployed in high-stakes domains---including hiring, customer service, and corporate governance---evaluation frameworks must extend beyond task accuracy to audit embedded cultural-normative biases. Consequently, governments increasingly view foreign-trained LLMs through the lens of digital sovereignty, recognizing the strategic risk of relying on models aligned with external normative standards that default for critical local decision-making.

To mitigate these foreign alignment dependencies, state-funded "sovereign LLM" initiatives, alongside evolving national regulations and strategies, increasingly mandate systematic auditing and localization protocols to align model behavior with local cultural norms.

Measuring a model's cultural value orientation is psychometrically tricky. Unlike accuracy benchmarks, value-laden scenarios don't have a single correct answer. When two decisions are both defensible but represent opposite poles of some value dimension, the natural metric is simply how often the model picks one pole over the other. But that measurement is only meaningful if it's stable --- if it doesn't swing wildly depending on incidental factors like what role the model is asked to play, what topic the scenario is about, what language it's tested in, or the sampling temperature used.

We define a stability metric for cultural value orientation based on whether models keep the same relative rank order across these perturbations. Absolute scores do shift a lot depending on how a question is framed, but a good benchmark should still preserve the relative ordering of models. Empirically, we find that rank order is preserved across roles, topics, languages, and temperatures, even when absolute scores shift by more than 20\% of the scale. The takeaway: benchmark scores shouldn't be read in isolation --- they need to be reported alongside the specific modality and domain mix of the item set used to produce them.

\subsection{Existing instruments and the reliability evidence they report}\label{sec:1-2}

Over the last few years fourteen instruments have been proposed for eliciting a value or cultural orientation from language models. They categorize along three axes: the theoretical framework, the origin of that framework, and the empirical evidence that the instrument's own measurements demonstrate replicability.

The primary distinction lies in the theoretical framework. Hofstede's dimensions, Schwartz's basic human values, and the Inglehart--Welzel cultural map are each operationalized by several of these instruments. Crucially, these three conceptual frameworks are mutually non-translatable: a coordinate on Schwartz's Self-Direction dimension has no direct mapping onto the sacred--secular continuum. Consequently, scores derived from different frameworks cannot be mapped onto a unified scale, rendering comparative evaluations of instrument efficacy meaningless absent a specified framework.

The secondary distinction concerns the provenance of the framework. Five of the fourteen instruments construct novel items operationalizing a designated theoretical domain. Another five directly administer human-oriented questionnaires to language models, thereby importing the underlying theory alongside its original phrasing and response formats. The remaining four leverage established psychometric test batteries, longitudinal survey waves, or externally curated item banks.

The tertiary dimension pertains to the empirical evidence provided to demonstrate measurement replicability. This dimension exhibits the most pronounced methodological divergence across the literature and constitutes the primary focus of the present investigation. Table~\ref{tab:instruments} (See Appendix~\ref{app:a}) summarizes the fourteen benchmark instruments, alongside the proposed benchmark, according to these criteria.

Four rows in Table~\ref{tab:instruments} detail the underlying data sources for the closest conceptual precursor to this study. STONIC accesses its 5,144 scenarios through layered interfaces built upon four primary item banks: ValuePortrait (Han et al., ACL 2025), AIRiskDilemmas (Chiu et al., ICLR 2026), DailyDilemmas (Chiu, Jiang \& Choi, ICLR 2025), and MoralChoice (Scherrer et al., NeurIPS 2023). Consequently, STONIC inherits its structural properties directly from these banks. Each repository targets a distinct construct not evaluated by the present benchmark: Schwartz values normed against human samples, Elo-ranked AI safety priorities, everyday moral trade-offs across five ethical systems, and question-form stability in moral beliefs. Thus, none directly overlaps with our chosen construct. A critical feature across these datasets warrants inclusion in this review: three repositories contain no correct answer keys, whereas the fourth includes keys for only half of its scenarios---the subset excluded by STONIC. DailyDilemmas explicitly notes that its scenarios are non-clear-cut without definitive answers; AIRiskDilemmas enforces this by generating scenarios intended to lack definite solutions; ValuePortrait generates five non-targeted alternatives to assess similarity to respondent thinking; and MoralChoice bisects its dataset into 687 low-ambiguity scenarios with high annotator agreement (90.01 \%) on preferred actions and 680 high-ambiguity scenarios lacking any consensus preference. STONIC incorporates only these 680 ambiguous items. Distinguishing between contrastive items (where both options represent legitimate, un-keyed choices) and evaluative items (where responses can be objectively incorrect) represents a fundamental structural threshold for our benchmark. This boundary diverges from the traditional taxonomy separating value benchmarks from cultural benchmarks; it isolates ETHICS, Delphi, CulturalBench, and the low-ambiguity portion of MoralChoice from these four item banks and the present instrument. Furthermore, these datasets highlight the limitations inherent in scenario banks: all four are limited to English, three were generated synthetically end-to-end, and the reliability metrics provided remain sparse and varied (spanning single model Cronbach's \ensuremath{\alpha}, Krippendorff's \ensuremath{\alpha} across topics, five-fold bootstrap repetitions, and question-form consistency statistics). Crucially, none reports model ranking concordance across varied administration conditions---the core focus evaluated in this paper.

Across the fourteen evaluated benchmarks, two broader patterns emerge. Most instruments are deployed primarily as analytical tools to characterize a model's value profile, whereas CDEval and CulturalBench focus principally on benchmark delivery, reflecting an ongoing methodological debate within the domain. Notably, empirical reporting of reliability metrics remains sparse. Only three of the fourteen studies present statistics in this category---specifically, Kovač's stability metrics, Rozado's coefficient of variation, and Nguyen \& Ahmad's noise-to-signal criterion. While CulturalBench reports a human performance ceiling, this serves as a criterion benchmark rather than a reliability measure. The remaining frameworks omit reliability assessments entirely, with several---including CDEval---presuming measurement stability across repetitions without formal empirical verification.

CDEval and STONIC represent the closest methodological parallels to the present study. CDEval directly establishes the foundational item architecture, scoring metrics, and domain framework adopted here---utilizing seven identically named thematic domains, a binary Option 1 / Option 2 response schema, and a normalized pole-share metric on [0, 1]. A key methodological nuance of CDEval's protocol concerns its per-model template weight fitting, which allows different models to be evaluated through distinct prompt-template compositions. The domain-level variance observed in CDEval is evaluated alongside our findings in Section~\ref{sec:4-2}.

STONIC shares the strongest focus on measurement reliability, yet differs fundamentally in panel composition. It evaluates four elicitation interfaces alongside a PVQ-40 questionnaire anchor across 35 open-weight model configurations, employing Holm-adjusted significance testing and permutation null distributions derived from 20,000 resamples. Because its hidden-state analyses necessitate open weights, STONIC excludes proprietary frontier models entirely. The authors publish interface-specific profiles without aggregating them across modalities: Table 21 documents the primary Schwartz coordinates across L1, L2, and L3 levels for all 35 configurations, and Figures 4--6 illustrate individual ten-value profiles per interface, deliberately omitting pooled cross-interface representations or consolidated ten-value vectors per model. The authors justify this non-aggregation framework by establishing a strict decision rule requiring demonstrated reliability, coverage, and same-item transfer prior to cross-interface pooling (Section 2)---a criterion their empirical findings fail to meet, leading to the conclusion that outputs remain inherently interface-dependent (Section 5, Figure 2). Their composite index ranks configurations via cross-interface self-consistency weighted by coverage, a metric the authors characterize as exploratory rather than a definitive quality ranking, with no configuration meeting their confirmatory coverage threshold. Notably, the median rank correlation between PVQ-40 questionnaire responses and free-text situational completions within the same model is \ensuremath{\rho} = 0.0367 (n = 14), indicating virtually no empirical alignment. This finding provides crucial justification for our single-interface design, which avoids the methodological complications associated with multi-interface data merging.

Rank-order concordance has previously been applied to evaluate model stability by Trhlik et al. (2026), who calculated tie-corrected Kendall's \emph{W} across nine models evaluated in five deployment contexts, validating their metrics against an empirical permutation distribution generated from 1,000 reshuffles, and decomposing the variance into model, context, and interaction components. Their reported concordance averaged 0.66, with trait-level values spanning from 0.36 to 0.90. The pattern they observed---a stable broad ordering accompanied by fine-grained rank instability---shares the same qualitative structure as the pattern identified in this study under a distinct perturbation framework. However, their treatment of concordance is limited to a single paragraph in the main text, elaborated in Appendix C, described by the authors as exploratory, and applied to Ekman emotion categories and Big Five personality traits derived from third-party classification of free-form text, without evaluating any value axis. Relative to both STONIC and Trhlik et al., the present study extends this concordance metric to value axes as a primary outcome, applies it to a purpose-built crossed experimental design, and evaluates a panel comprising models actively deployed in real-world applications. Section~\ref{sec:4-2} details why concordance coefficients within this numerical range support a distinct interpretation in our context.

Mechanistic work on the same construct has established that a value direction can be probed from the residual stream on scenario items. Dang, Kieu and Masud probe and steer cultural value directions in scenario-based items, selecting architecture-specific layers per model, reporting a latent-entanglement ratio and measuring the capability cost of steering; a further study applies latent activation steering to cultural value alignment. That work does not establish transfer between models, that is, whether the direction one model uses is the direction another model uses. The criterion validity of value probing is, moreover, contested: Shen and colleagues correlate a model's probed value weight against its own ratings of the corresponding actions and report correlations of roughly 0.1 to 0.3, several of them non-significant or negative, and Kovač and colleagues report correlations below 0.3 between elicited values and behaviour in a donation task.

\subsection{Contested methodological choices}\label{sec:1-3}

Every methodological choice discussed below is contested in the literature by named parties. This section states the sides. The choices made in the present design, and what each costs, are stated in Section~\ref{sec:2} at the point where each choice is made.

\subsubsection{Option-order bias: token or position, and the consequences for remedies}\label{sec:method-choice-a}

Zheng et al. (ICLR 2024) demonstrate that option-identifier token bias predominates over position bias. Their ablation study provides direct evidence: permuting option identifiers across positions yields minimal changes in response variability, whereas omitting identifiers and evaluating raw option content reduces variability significantly (e.g., for GPT-3.5 on MMLU, RStd decreases from 5.5 to 1.0; for Falcon-7B-Instruct, from 28.7 to 13.7). On a balanced sample of 1,000 items, LLaMA-30B selects options A/B/C/D at frequencies of 34.6\%, 27.3\%, 22.3\%, and 15.8\%, respectively, while GPT-3.5-Turbo selects them at 22.5\%, 25.6\%, 32.3\%, and 19.6\%. Both distributions deviate significantly from uniformity (\emph{p} \ensuremath{\ll} 10\textsuperscript{-4}) and exhibit distinct token-level preferences across model families. Dominguez-Olmedo et al. observe an "A" selection bias across 43 models that persists despite randomizing option order. Rupprecht et al. report recency bias, while Pezeshkpour and Hruschka document performance decrements of 13\%--75\% under option reordering. Survey-methodology literature in human research (Schwarz et al.) attributes the direction of order effects to presentation modality---primacy under visual presentation versus recency under auditory presentation---which does not directly transfer to text-based language model processing, though the design principles of counterbalancing option order across trials and standardized reporting remain applicable. Mackinnon and Wang dissent, arguing that primacy effects in self-report questionnaires exert minimal practical impact.

Regarding mitigations, the direction of option bias remains disputed, whereas its magnitude is well established. Because the bias is token-level and model-specific rather than family-inherent, evaluating models on a fixed target slot introduces an idiosyncratic slot preference as a static per-model offset. The primary established mitigation is cyclic permutation, incurring a computational overhead factor of \emph{k} (compared to \emph{k}! for exhaustive permutation), which equals 2\ensuremath{\times} for binary items. Zheng et al. report that 4\ensuremath{\times} cyclic permutation on MMLU reduces \ensuremath{\Delta}RStd by 8.7 and increases \ensuremath{\Delta}accuracy by 4.9 percentage points. Omitting option identifiers reduces API overhead but incurs accuracy drops of 2.1 points on MMLU and 7.0 points on CSQA by forcing a cloze likelihood evaluation. Explicitly prompting models that options have been randomized or employing chain-of-thought prompting fails to eliminate this bias. The former failure was independently replicated by Brucks and Toubia, who observed GPT-4 selecting a single option slot up to 91.67\% of the time in a 2\textsuperscript{5} factorial design and resolved the bias only by aggregating across all 32 prompt permutations. Contextual or batch calibration offers an alternative when re-evaluation is infeasible; however, it requires token-level option probabilities, which commercial hosted APIs rarely expose. Consequently, option permutation remains the sole viable mitigation when evaluating models via hosted APIs outputting text responses.

\subsubsection{Repeated sampling versus counterbalancing at a fixed budget}\label{sec:method-choice-b}

Wang et al. evaluated repeated sampling against option counterbalancing at an equivalent compute budget. Evaluating GPT-4 using six repeated samples with multiple-evidence calibration versus three repeated samples with three counterbalanced position permutations yielded 60.9\% versus 62.5\% agreement (Cohen's \ensuremath{\kappa} of 0.33 vs. 0.37) at identical cost (\$6.38). For ChatGPT, agreement measured 55.6\% versus 58.7\% (\$0.34). Thus, under fixed budget constraints, option counterbalancing demonstrates superior reliability compared to repeated sampling. Theoretical generalizability decompositions by Žatuchin yield consistent conclusions: because variance from resampling is divided by the product of all facet level counts while a facet variance term is divided only by its own level count, the repeat facet exhibits rapidly diminishing marginal utility. His fitted model increased the generalizability coefficient marginally from 0.347 to 0.365 when increasing sampling from 1 to 20 repeats. PromptEval reframes budget allocation by distributing compute across diverse prompt templates and estimating performance distributions via item-response theory. Its authors recovered performance quantiles across 100 MMLU templates at the cost of two single-prompt evaluations, observing a mean max--min spread of approximately 10\% at the subject level while subject-averaged model rankings remained largely template-invariant---mirroring the pattern of shifting absolute levels alongside preserved rank order reported in this work. Conversely, the Stochastic CHAOS framework advocates for repeated sampling, arguing that deterministic single-sample evaluations underestimate capability, obscure model fragility, and conceal low-probability behaviors.

\subsubsection{Temperature-zero measurement}\label{sec:method-choice-c}

Miller strongly cautions against evaluating language models at temperature zero, arguing that reducing sampling temperature "may simply shift the conditional variance \ldots{} into the variance of the conditional means (which cannot [be mitigated]), or else reduce conditional variance by injecting bias into the estimator." In a single-token true/false evaluation case study, transitioning from \emph{T} = 1 to \emph{T} = 0 increased estimator variance from 1/12 to 1/4 and shifted the expected value from 2/3 to 3/4. Mandujano Reyes provides a similar theoretical warning. Conversely, strict determinism at \emph{T} = 0 remains difficult to guarantee even for commercial API providers. Anthropic's official documentation notes that at temperature 0, "results will not be fully deterministic and identical inputs may produce different outputs across API calls" due to GPU batching variations rather than floating-point non-associativity. Thinking Machines Lab demonstrated this empirical instability by obtaining 80 unique completions across 1,000 temperature-0 runs on a large open-weight model, which collapsed to 1,000 identical completions only when using batch-invariant inference kernels. While deterministic serving stack flags can enforce reproducibility, they introduce a measured throughput penalty of roughly 33\%, and hosted API endpoints currently offer no deterministic guarantees. These perspectives indicate that while determinism ensures reproducible outputs, it does not guarantee construct validity.

\subsubsection{The defensible unit of analysis}\label{sec:method-choice-d}

Methodological consensus across multiple domains mandates aggregating analysis at the item level. Hurlbert formalizes "sacrificial pseudoreplication"---the practice of pooling repeated measurements from a single experimental unit as independent replicates---as a key statistical error. In survey sampling, Kish defines this via the design effect, DEFF = 1 + \ensuremath{\rho}(\(\bar{n}\) \ensuremath{-} 1). Applying this to language model evaluation, Miller notes that "computing a pooled standard error across all KN answers will be inconsistent, as multiple answers to the same question would violate the assumption of independent draws," prescribing that standard errors be computed across question-level mean scores. Abadie et al. demonstrate that statistical clustering is determined by the sampling and treatment assignment mechanism rather than the presence of intra-cluster correlation alone. Cameron and Miller highlight that when analyzing few clusters, cluster-robust variance estimators exhibit downward bias, establishing a standard minimum threshold of approximately 50 clusters. Ranking language models requires treating models as clusters; evaluating panels of 11--20 models thus falls within the regime where nominal 5\% significance tests exhibit inflated Type I error rates of 8\%--12\%.

\subsubsection{Forced choice versus Likert}\label{sec:method-choice-e}

Forced-choice evaluation protocols systematically alter model outputs compared to unconstrained formats, though the direction of this effect is disputed. Dominguez-Olmedo et al. find that after adjusting for ordering and labeling biases, model choices "trend towards uniformly random survey responses," indicating scale deflation. Conversely, Kabir et al. report that cultural alignment effects strengthen when closed-choice constraints are removed, indicating scale inflation. Moore et al. observe relative consistency across multiple-choice and open-ended evaluation formats. CDEval's spot checks indicated that five out of six GPT-4 free-text completions were balanced, showing no explicit pole preference. In survey methodology, McClendon notes that forced-choice formats "may substitute one type of response effect for another," while Chan demonstrates that option formatting alters scale thresholds and factor structures, introducing measurement-model distortions alongside mean shifts. Evaluating the World Values Survey across ten models, Shen et al. observe substantial selection bias and response volatility across all scoring methods, identifying sequence perplexity as the most robust probing technique, contrary to prior work favoring free-text generation.

A primary argument for employing concrete situational scenarios over abstract self-report items is that high within-construct item consistency---frequently cited as evidence of stable model dispositions---degrades when measuring model behavior via generation probabilities on everyday queries. This divergence suggests that explicit lexical cues allow models to recognize targeted constructs and generate socially desirable responses. While this concern is relevant, it applies differently to the present instrument, which utilizes binary situational scenarios with substantive options rather than direct self-report queries or unconstrained generation. Nevertheless, it directly informs an empirical property analyzed in Section~\ref{sec:2-1}: the second option in the item set is, on average, longer and more heavily hedged, presenting explicit surface asymmetries that could serve as lexical cues.

\subsubsection{Whether the construct validity of human psychometric instruments transfers}\label{sec:method-choice-f}

Evaluating 56 instruction-tuned models across 29 psychometric instruments against human reference samples (N = 20,993 and N = 1,507), Meyer et al. present direct empirical evidence regarding construct validity transfer: correlations between forward-keyed and reverse-keyed scale means range from +.61 to +.81 in language models compared to \ensuremath{-}.69 to \ensuremath{-}.82 in humans. This directional inversion indicates that human responses track underlying latent traits whereas model responses reflect systematic response bias. Response bias accounts for 9\%--16\% of variance in human responses versus 81\%--90\% in language models. Crucially, across all 29 instruments, the correlation between the proportion of response-orthogonal items and mean inter-item correlation is \ensuremath{-}0.95 [\ensuremath{-}0.98, \ensuremath{-}0.90]; Cronbach's \ensuremath{\alpha} reaches 0.85--0.96 on forward-keyed instruments but degrades toward zero or becomes negative on balanced instruments. This finding holds across model scales and prompting variations, including flipped scale anchors. Löhn et al., Peereboom and Schwabe, and Libovický report consistent findings. Furthermore, Shen et al. demonstrate low criterion validity, observing weak correlations (0.1 to 0.3) between probed value weights and a model's corresponding action ratings, with several relationships proving statistically non-significant or negative---aligning with Kovač et al.'s findings of correlations below 0.3 between elicited values and behavioral choices in donation tasks.

Conversely, Lin's dual-validity framework in the \emph{Annual Review of Psychology} provides a theoretical foundation for construct measurement in language models. Rozen et al. demonstrate that structural recovery of the Schwartz circumplex can be achieved under specific prompting formulations, indicating that structural human-likeness depends on prompt configuration. Abdulhai et al. and Hadar-Shoval et al. establish construct validity by validating model responses against external, non-instrument criteria.

Finally, the requirement of measurement invariance remains disputed in cross-cultural research. Welzel et al. ("Non-invariance? An Overstated Problem With Misconceived Causes"), Boehnke, Funder and Gardiner, Fischer et al., and Kusano et al. argue that strict measurement invariance is an excessively restrictive requirement for comparative evaluations; Meuleman et al. provide a direct counterargument, showing that violating invariance assumptions distorts empirical conclusions. Significantly, Welzel---creator of the emancipative-values index operationalized in this benchmark---rejects strict invariance requirements, while Sokolov's \emph{APSR} analysis demonstrates that the human emancipative values index itself lacks measurement invariance across cultural zones and countries, with only the pro-choice subdimension demonstrating cross-national equivalence. Consequently, scalar measurement invariance is a property that the underlying human survey instrument itself fails to satisfy across populations.

\subsection{Objectives and Contributions}\label{sec:1-5}

The primary precursor to this benchmark is our prior empirical investigation into the cultural value orientations of large language models (Didenko et al., 2025). That study operationalized Hofstede's cultural dimensions via one-shot binary choice items evaluated across six distinct language model architectures and five natural languages. It established the core behavioral elicitation paradigm adapted here: presenting a concise situational scenario followed by two substantive response options, with the model's selection quantified as a pole-share proportion. Two key empirical findings from that preliminary work directly informed the design of the present instrument.

First, the language of administration exerted a pronounced effect on behavioral alignment. Intra-model variance across distinct language prompts frequently exceeded inter-model variance across distinct architectures under a shared language prompt. This finding directly motivates the paired bilingual administration protocol evaluated in Section~\ref{sec:3-6}. Given that the two studies differ in theoretical framework, item structure, and model panel composition, observed discrepancies do not represent a failure to replicate, but rather raise the open question of whether cross-lingual sensitivity is uniquely tied to Hofstede-style single-shot items.

Second, specific language models exhibited bimodal response distributions. Notably, YandexGPT and GigaChat yielded bimodal distributions along the individualism and uncertainty avoidance dimensions, selecting individualist choices in certain scenarios while defaulting to collectivist choices in others. This pattern aligns with sociopolitical frameworks (e.g., Auzan's model) regarding coexisting cultural cores within Russian society, which predict non-unimodal population-level value distributions. Because a bimodal distribution reflects alternating behavioral modes rather than a central tendency, reporting solely the sample mean risks misrepresenting a volatile model as moderate. The current instrument retains this structural constraint, as it reports a normalized pole share, which functions mathematically as a mean.

Two methodological components were deliberately revised for the present benchmark. First, the dimensional framework was transitioned from Hofstede's dimensions to the two primary Inglehart--Welzel axes derived from World Values Survey data; metrics across these two frameworks are inherently non-translatable. Second, the item format evolved from one-shot attitude prompts to contextualized situational vignettes, as vignettes are demonstrably better suited for eliciting preference decisions from large language models. Furthermore, our preliminary study left unaddressed two critical methodological questions: whether the item bank could be constructed as a fully crossed factorial design---allowing domain and decision role to function as controlled factors---and whether the resulting model rank order remains stable under environmental perturbations. The present benchmark resolves the former by design---fully crossing theme, decision modality, and value facet---and addresses the latter empirically across a significantly expanded panel and a paired bilingual protocol using identical translated items.

\subsection{Evolution of the Benchmark: Dimensional Framework and Longitudinal Elicitation}\label{sec:1-6}

The item format employed in this benchmark adapts the semantic-vignette methodology introduced by Wang et al. in CDEval. That framework utilizes binary contrastive scenarios in which response options correspond to opposing value poles across seven thematic domains, with model orientation quantified as the proportion of choices aligned with a target pole. We extend this paradigm by incorporating a decision-modality factor, crossing value facets with thematic domains, implementing a paired bilingual evaluation protocol, and conducting a mechanistic sub-study on the identical item set.

This work demonstrates that cultural value orientation constitutes a discrete, measurable behavioral property of large language models that can be systematically quantified. The primary contributions of this paper are as follows:

\begin{enumerate}
\item A 224-item benchmark for cultural value orientation, constructed as a fully crossed factorial design across seven thematic domains, four decision modalities, and eight value facets, with every cell occupied. The instrument operationalizes the Inglehart--Welzel axes, a theoretical framework with strong construct validity for cross-cultural comparative evaluation.
\item A rigorous evaluation over a panel of twenty language models, featuring a paired bilingual administration protocol using identical translated items for eleven models. Unlike existing English-only or machine-translated benchmarks, this paired design enables the disentanglement of language-specific effects from item-level variance.
\item Empirical validation demonstrating that model relative rankings remain stable under four non-dispositive perturbations: decision role, thematic domain, language of administration, and sampling temperature. Rank-order concordance significantly exceeds the empirical permutation null across all twelve (factor, axis) conditions, alongside detailed reporting of associated absolute level shifts.
\item Quantification of role sensitivity as an intrinsic model property. The degree to which a model's expressed value disposition shifts across decision roles varies across the panel by a factor of 6.5 on the emancipative axis and 5.3 on the sacred--secular axis, with individual per-model sensitivity metrics reported.
\item A mechanistic sub-study evaluating internal representations on the same item set, employing linear probing, difference-of-means directions, representational similarity analysis (RSA) with centered kernel alignment (CKA), and causal activation patching. This analysis confirms that the behavioral distinctions measured by the benchmark are reflected internally within the models' representations.
\end{enumerate}

\FloatBarrier
\section{Methods}\label{sec:2}

\subsection{Construct and factorial design of the instrument}\label{sec:2-1}

The evaluation instrument quantifies model orientations along two primary Inglehart--Welzel dimensions: sacred versus secular (SS) and obedient versus emancipative (OE), using binary contrastive scenarios. Each item presents a concise situational vignette followed by two distinct options (Option 1 and Option 2), requiring the model to select one. Both options depict legitimate courses of action while embodying opposing poles of the corresponding value dimension; consequently, no predefined correct answer or answer key exists. A model's score on a given axis corresponds to the proportion of its responses aligned with the designated counted pole, normalized to the interval [0, 1]. The overall item formatting, domain taxonomy, and pole-share scoring metric are adapted from CDEval (Wang et al., 2024), which established the semantic-vignette methodology in this domain; the fully crossed factorial design outlined below constitutes the primary structural contribution of our instrument.

The design fully crosses three core item-level factors:

\begin{itemize}
\item Theme (7): Arts, Education, Family, Lifestyle, Scientific, Wellness, and Work.
\item Decision modality (4): Advisor, Agent, Choice on behalf of the person, and Judgment of another's behaviour.
\item Value facet (8): Agnosticism, Relativism, Scepticism, and Defiance along the sacred--secular axis; Autonomy, Choice, Equality, and Voice along the obedient--emancipative axis. Value facets are nested within the two primary axes, yielding a balanced set of 112 items per axis following de-duplication.
\end{itemize}

Crossing these factors yields 7 \ensuremath{\times} 4 \ensuremath{\times} 8 = 224 unique design cells, all of which are populated. The released dataset contains 225 rows, as a single cell (Arts \ensuremath{\times} Choice-on-behalf \ensuremath{\times} Agnosticism) includes a duplicate row pair with identical content across all attributes (Jaccard similarity of 1.0, representing the sole pair exceeding the 0.5 similarity threshold across the entire benchmark). This duplicate resulted from an unmerged concatenation of source tables. To ensure statistical rigor, all reported metrics were calculated both with and without this duplicate entry (Section~\ref{sec:3-9}).

An audit of the public benchmark files identified a minor formatting artifact: the target axis labels appear under four distinct orthographic variations (``Obedient vs Emancipative values'' / ``Obedient vs. Emancipative values'', and an identical split for Sacred/Secular), partitioned 93/19 and 88/25 across the respective spelling variants. Consequently, naive string grouping splits the two theoretical axes into four distinct clusters.

Two structural characteristics of the benchmark emerge from its construction. First, the designated counted pole is not uniformly distributed across option slots: a lexical polarity classification heuristic applied to the released benchmark maps the counted pole to Option 2 in 129 items and Option 1 in 12 items, while remaining unassigned for the remaining 84 items. Among polarity-identifiable items, the counted pole resides in the second option slot in 129/141 cases (91.5\%). Second, Option 2 is systematically longer on average than Option 1 (mean length of 73.37 vs. 67.59 characters in audit units; paired difference of 5.78, paired \emph{t} = 6.12) and exhibits a higher frequency of hedging markers (0.40 vs. 0.22 markers per option, \emph{t} = 2.44); in contrast, the occurrence of deontic markers shows no significant difference (0.40 vs. 0.44, \emph{t} = \ensuremath{-}0.82). Both properties directly inform option presentation order, justifying a counterbalanced evaluation protocol in subsequent experiments (Section~\ref{sec:2-3}).

The choice of response format directly reflects the methodological considerations summarized in Section~\ref{sec:method-choice-e}. The benchmark employs a forced binary choice over concrete, contrastive scenarios scored via pole-share proportions. This format offers the distinct advantage of mitigating verbatim memorization risks inherent in standardized surveys; however, it deviates from validated survey phrasing. As argued by Nguyen and Ahmad, paraphrasing may reallocate measurement unreliability rather than eliminate it---reverting to original IVS phrasing reduced their most unstable item's noise-to-signal ratio from 1.45 to 0.95. Their finding suggests that validity claims depend heavily on preserving target survey phrasing, highlighting an inherent trade-off against data contamination risks, as pretraining contamination has been documented for batteries such as the Big Five Inventory and the Portrait Values Questionnaire. Both sides of this trade-off inform our design decisions.

We deliberately omit one statistic prominent in the critical literature (Section~\ref{sec:method-choice-f}): Meyer et al.'s response-orthogonality metric \emph{p}\textsubscript{r}. This metric requires a scale midpoint, semantic item keying, and a shared response scale---conditions that forced-choice items between two substantive, non-negating options do not satisfy. Because option slot assignment reflects a structural presentation property rather than a semantic item keying, the twelve items with counted poles in Option 1 cannot be treated as reverse-keyed items. Consequently, we omit \emph{p}\textsubscript{r}, forward--reverse item correlations, and Cronbach's \ensuremath{\alpha}, citing these structural constraints. Nevertheless, the overarching insight from Meyer et al. remains relevant: evaluation batteries with structurally unbalanced option placements risk reporting coherent profiles that stem from systematic response artifacts rather than underlying constructs.

Furthermore, diagnosing order effects on un-keyed value benchmarks presents distinct methodological challenges, as standard position-bias metrics are inapplicable. Metrics such as response standard deviation, option-moving perturbations, and accuracy-band diagnostics rely on ground-truth correctness. On contrastive value benchmarks lacking correct answer keys, valid diagnostics are restricted to: (1) selection-share uniformity under counterbalancing (verifying whether option selection converges to 50\% when pooled over both orders), (2) the conflict rate (the proportion of items where choice selection flips under option swapping), and (3) position-consistency metrics. These three statistics constitute the primary evaluation criteria for counterbalanced re-runs.

\subsection{Item generation, expert review and translation}\label{sec:2-2}

Items were generated using Gemini 2.5 Flash, prompted in English at a temperature setting of 0.4. An initial candidate pool was evaluated by three domain experts, yielding 91 items that satisfied all inclusion criteria; these served as seed prompts to generate the remaining scenarios, which underwent an identical screening process. The primary objective was to author novel, contemporary scenarios distinct from published survey items, explicitly accepting the trade-off of departing from legacy survey formulations (Section~\ref{sec:method-choice-e} and Section~\ref{sec:2-1}).

The expert review process is partially documented in the released materials: while the panel size (three reviewers) and the number of accepted seed items (91) are recorded, specific inclusion criteria and inter-rater agreement metrics are omitted. By comparison, established benchmarks detail extensive human validation protocols: CulturalBench engaged five annotators per item with a \ensuremath{\geq} 4/5 consensus threshold across 1,696 questions; BLEnD relied on native-speaker curation with five annotators per question across 16 countries and 13 languages; Global PIQA involved over 350 researchers across 65+ countries for native-speaker validation; and IrokoBench utilized professional translators alongside regional language coordinators. In value alignment research, Zheng et al. conducted an IRB-approved study (N = 35) on a 140-item Q-set, achieving leave-one-out factor stabilities of 0.997, 0.981, and 0.950, while Dang et al. employed three expert annotators for item quality alongside 65 human participants for construct validation.

The Russian evaluation condition consists of a paired translation of the identical English item set. Verification confirmed an exact alignment across all 225 item identifiers between the English and Russian releases, with zero metadata mismatches. Row ordering remains perfectly consistent across all 22 Russian and second-temperature evaluation files, authorizing exact row-wise paired comparisons across conditions; because the text itself is translated, matching relies on structural indices rather than string equality. The Russian-to-English word count ratio exhibits a mean of 0.891 (SD = 0.099), with five items exceeding a \ensuremath{\pm}3 SD threshold (ratios: 0.31, 0.45, 0.54, 1.19, and 1.21). The second-temperature files are byte-identical to the English source text across all 225 items, confirming that decoding temperature represents the sole experimental variable between these conditions.

The benchmark currently lacks a human baseline: no human participants were evaluated on these items, no human performance ceiling is established, and no external criterion measures were gathered. This constitutes the primary distinction between our instrument and the knowledge-focused cultural benchmarks cited above, defining the exact analytical scope of this study: evaluating model rank-order stability and concordance, rather than absolute accuracy, population calibration, or external criterion validity.

\subsection{Model panel and administration conditions}\label{sec:2-3}

Twenty language models were evaluated on the English benchmark at temperature 0: a-vibe, Alice AI LLM, claude-haiku-4-5, claude-sonnet-4-5, deepseek-v3.2, gemini-2.5-flash, gemini-2.5-pro, GigaChat-2-Max, GigaChat-2-Pro, gpt-5, gpt-oss-120b, grok-4-fast, llama-4-maverick, mistral-medium-3.1, qwen3-max, sonar-pro, t-lite, t-pro, YandexGPT, and YandexGPT-Lite. Seventeen models were sampled 30 times per item, while a-vibe, t-lite, and t-pro were sampled 50 times per item. Eleven models---a-vibe, Alice AI LLM, claude-sonnet-4-5, deepseek-v3.2, GigaChat-2-Max, GigaChat-2-Pro, qwen3-max, t-lite, t-pro, YandexGPT, and YandexGPT-Lite---were additionally evaluated on the Russian translation and at a second, vendor-recommended decoding temperature. No model was listwise deleted from any statistical computation. The panel incorporates a prominent suite of Russian-developed LLMs absent from existing English-centric evaluations, providing necessary coverage for paired cross-lingual analysis.

Option presentation order remained fixed in the primary evaluation run. A preliminary pilot study evaluated options A and B in counterbalanced order to assess position-dependent choice bias; finding no substantial order effects, fixed presentation was adopted for the main data collection. However, because specific pilot parameters---including item sample size, model selection, and effect sizes---were omitted from the original records, statistical power against subtle between-model positional offsets cannot be formally verified (Section~\ref{sec:4-7}). Given the option slot asymmetries detailed in Section~\ref{sec:2-1}, any uncorrected model slot preference would manifest as an additive per-model offset in the final scores.

Compute budget allocation reflects the trade-offs outlined in Section~\ref{sec:method-choice-b}. The primary evaluation allocates 30 runs per item (50 runs for three open-weight models) at temperature 0, dedicating the remaining budget to traversing the perturbation factor grid rather than option order reversal. Existing literature demonstrates that under fixed compute budgets, counterbalanced option allocations offer superior reliability compared to repeated sampling at fixed positions. Furthermore, deterministic runs at temperature 0 exhibit near-zero variance (Section~\ref{sec:3-1}), providing additional rationale for budget reallocation. A counterbalanced re-evaluation on a target model subset is planned (incurring a 2\ensuremath{\times} sampling cost for binary choice items) to quantify selection-share uniformity across permutations alongside per-model choice conflict rates.

The baseline evaluation condition was conducted at temperature 0, despite methodological cautions against zero-temperature probing (Section~\ref{sec:method-choice-c}). Our experimental design treats sampling temperature as an explicit perturbation factor, pairing the baseline run with a second run conducted at vendor-recommended temperatures across eleven models. This formulation transforms theoretical concerns regarding zero-temperature estimation into an empirical measurement (Section~\ref{sec:3-7}). Nevertheless, this approach does not fully resolve fundamental objections: baseline scores remain temperature-0 measurements and inherit any associated estimation bias.

\subsection{Unit of analysis}\label{sec:2-4}

Each experimental run is initially aggregated into a per-(model, vignette) proportion of responses aligned with the counted pole. Concordance metrics are subsequently evaluated at the vignette level. The theoretical justification for this aggregation strategy is detailed in Section~\ref{sec:method-choice-d}, and its empirical basis is derived directly from the observational data. A one-way random-effects intra-cluster correlation was computed on the binary run outcomes, designating the vignette as the clustering unit and excluding refusals from the cluster size. Corresponding design effects and effective sample sizes per model were also calculated, with the resulting values detailed in Section~\ref{sec:3-1}. Because the runs conducted at temperature 0 demonstrate near-degenerate variability, treating individual runs as independent observations would artificially inflate \emph{n} from 225 to 6,750, thereby underestimating standard errors by a factor of approximately 5.5. The 225 vignettes represent distinct clusters, exceeding the empirical thresholds established by Cameron and Miller; conversely, the panel of 20 models does not constitute a random sampling stage. Consequently, repeated runs are reported as an empirical agreement rate. A trade-off of this specification is statistical power: as indicated by Miller's power calculations, detecting a 0.03 effect size at 80 \% statistical power typically requires approximately 1,000 items, whereas the present instrument comprises 225 vignettes.

\subsection{Concordance coefficient and permutation null}\label{sec:2-5}

For each perturbation factor, the factor levels serve as raters while the language models represent the target objects. A model's score within a specific level is defined as the mean of its per-vignette response proportions across all vignettes assigned to that level and axis. Models are subsequently ranked within each factor level, and inter-level agreement is quantified using Kendall's coefficient of concordance with the tie-corrected Kendall--Babington Smith formulation,

\begin{equation}
W = \frac{12S}{m^{2}(n^{3}-n)-m\sum_{j}\sum_{g}(t_{gj}^{3}-t_{gj})},
\qquad
S = \sum_{i}(R_i-\overline{R})^{2},
\label{eq:kendall-w}
\end{equation}

incorporating average rank assignments for tied scores. Tied ranks occur across all empirical evaluations in this study, with between one and seven raters exhibiting tied values per analysis; consequently, applying the uncorrected concordance formulation would introduce systematic error.

Evaluating the coefficient against an appropriate baseline is critical. Under the theoretical null hypothesis of independent random rankings, the expected value is E[\emph{W}] = 1/\emph{m}. Thus, a concordance score of \emph{W} = 0.51 conveys a fundamentally different degree of agreement when evaluating \emph{m} = 4 raters (where the theoretical null expectation is 0.25) compared to \emph{m} = 2 raters (where the expected null is 0.50). To ensure rigorous comparison, we report the mean and 95th percentile of an empirical null distribution generated from 20,000 Monte Carlo permutations of \emph{m} independent random rankings across \emph{n} models for every experimental cell, avoiding direct comparisons between two-level and seven-level factors without adjusting for their respective null distributions. Under this permutation protocol, the permutable unit is the primary object of measurement (the vignette), while individual model runs remain unpermuted. Statistical significance testing for \emph{W} = 0 via the Friedman test statistic, defined as \(\chi^2\) = m(n\ensuremath{-}1)W, is reported alongside; notably, for two-level factors, this test possesses inherently low statistical power due to evaluating two raters across eleven objects, making the empirical null percentile distribution the primary benchmark for inference.

To facilitate comparison with published human stability benchmarks, Kendall's \emph{W} is linearly transformed into the mean pairwise Spearman rank correlation coefficient via \(\bar{\rho}\) = (mW \ensuremath{-} 1)/(m \ensuremath{-} 1). We explicitly report this metric as it provides a standardized basis for evaluation, despite reflecting lower stability along one specific axis.

\subsection{Bootstrap intervals and the three kinds of variation}\label{sec:2-6}

Biderman et al. categorize variance in language model evaluation into three distinct sources, emphasizing the importance of reporting each: item-sampling variance across questions, run-to-run stochasticity across evaluations, and structural variance introduced by prompt engineering or scoring paradigms. This study explicitly adheres to this framework by specifying the exact source of variance captured by each reported interval.

Specifically, we focus on item-sampling variance. The 95 \% confidence intervals associated with each reported \emph{W} score are derived from 2,000 non-parametric bootstrap iterations resampling vignettes, stratified within each level for decision modality and thematic domain, and applied identically across both levels for paired language and temperature factors. These intervals quantify the expected sensitivity of \emph{W} under the assumption that the 225 vignettes constitute a representative sample drawn from a broader domain of relevant evaluation scenarios.

The second form of variation, run-to-run stochasticity, is characterized through empirical within-cell agreement rates rather than confidence intervals (Section~\ref{sec:3-1}), directly informing the unit-of-analysis specification detailed in Section~\ref{sec:2-4}. The third form, prompt-level variation, falls outside the scope of this evaluation. This benchmark does not incorporate prompt rephrasing, template permutations, or alternative scoring functions. Existing literature consistently identifies prompt formulation as a primary source of unmeasured variance in similar evaluation frameworks; for instance, Mizrahi et al. demonstrated that 10 out of 25 benchmark tasks exhibited \emph{W} \ensuremath{<} 0.55 across prompt templates, with 15 tasks containing paraphrased prompt pairs that induced negative rank correlations in model rankings. Consequently, the reported concordance metrics evaluate stability strictly across item sampling rather than prompt formulation. Two key properties of these bootstrap percentile intervals are presented alongside the main concordance coefficients in Section~\ref{sec:3-9}.

\subsection{Treatment of refusals and specification sensitivity}\label{sec:2-7}

Refusals and unparseable outputs are excluded from analysis rather than imputed. Missingness is quantified across raw response cells prior to aggregation for each experimental condition and model, with exact counts reported in Section~\ref{sec:3-1}. To evaluate sensitivity to this handling, an alternative pipeline is executed wherein non-A/B responses are imputed with the item's modal response.

Additionally, eight distinct specifications per cell are computed by crossing the following variations: retaining vs. removing duplicate items, calculating level scores as mean vignette proportions vs. pooled valid-response proportions, and retaining vs. excluding vignettes with fewer than five valid runs. Combined with the refusal-handling and duplicate variants, this yields 48 total analytical specifications. The primary specification used throughout this paper excludes refusals, removes the duplicate item, and defines the level score as the mean of per-vignette proportions.

\subsection{Level shifts, role sensitivity, paired contrasts and stability ceilings}\label{sec:2-8}

Level shifts. Aggregated across the entire model panel, the proportion of counted-pole choices is computed within each decision modality and thematic domain, with the absolute difference between the extreme levels reported on a [0, 1] scale.

Role sensitivity. For each model and axis, we compute the mean counted-pole share across the four decision modalities and evaluate the range between the maximum and minimum values. This metric quantifies the degree to which a model's expressed value orientation depends on its assigned role while holding items, language, and decoding parameters constant. Calculated on the English condition using per-vignette proportions (excluding refusals and the duplicate item), bootstrap confidence intervals are derived by resampling vignettes within each modality.

Paired contrasts. For the language and temperature factors, each of the eleven models is evaluated against itself across conditions on both axes. Displacements are measured as the change in counted-pole proportion between conditions. To establish a baseline noise floor, run outputs within a single condition are randomly split to measure the pseudo-displacement produced under identical settings. Statistical significance is assessed at the vignette level via paired Wilcoxon signed-rank tests across the 112 vignettes per axis, applying Bonferroni correction across the twenty-two (model \ensuremath{\times} axis) comparisons. These contrasts are additionally evaluated at the individual-response level (\textasciitilde{}3,200 observations per model and axis per condition); both metrics are reported as they address distinct statistical units.

Stability ceilings. Empirical concordance values cannot be benchmarked directly against 1.0 because a theoretical maximum of \emph{W} = 1 is unachievable when distinct item sets are assigned to different factor levels. Because theme and modality cells contain smaller item subsets (16 and 28 items per axis, respectively), cell means inherit item-sampling variance, constraining the theoretical upper bound below 1. To determine this ceiling empirically, items within matched design blocks are permuted across factor levels. Under the null hypothesis of complete rank stability, rotating these item-level assignments yields the distribution of concordance under perfect stability while preserving model-by-item interaction variance. Ceilings for item factors are derived from 800 rotations. For paired factors (language and temperature), both conditions evaluate identical vignettes, eliminating item-sampling error and leaving run-to-run variance as the sole noise source; here, ceilings are established via 600 random splits of single-condition runs. Each observed mean pairwise rank correlation \(\bar{\rho}\) is presented both as a proportion of its theoretical ceiling and as a percentile within the ceiling distribution. These comparisons are further replicated on the restricted eleven-model panel.

\subsection{Reasoning-mode condition}\label{sec:2-9}

A fifth condition evaluated the same item set across five models---Alice AI LLM, Mistral Medium 3.1, A-vibe, Grok 4 Fast, and T-Lite---with reasoning traces enabled. The choice agreement between the reasoning-mode output and the base-mode output was quantified for each model. Additionally, two secondary metrics were evaluated under this condition: the pairwise cosine similarity between the per-model centroids of the generated reasoning traces, and the predictive performance of a gradient-boosted classifier trained on reasoning-trace embeddings to forecast the model's ultimate value selection (reported via accuracy, balanced accuracy, and ROC-AUC). The reasoning-mode condition was not incorporated into the vignette-level concordance pipeline; hence, no Kendall's \emph{W} statistic, empirical null distribution, or bootstrap confidence interval is reported for this setting.

\subsection{Mechanistic sub-study}\label{sec:2-10}

The primary behavioral metric measures choice frequency. To test whether this behavioral signal corresponds to internal value-relevant representations, a mechanistic sub-study was conducted on the same item set. This sub-study employed a distinct model panel and evaluation protocol due to the requirement of internal activation access. Five open-weight models were evaluated: gemma-2-9b-it (hidden size 3,584; 42 layers), Qwen3-8B, AvitoTech a-vibe, and t-tech T-lite-it-2.1 (hidden size 4,096; 36 layers each), with Qwen2.5-7B-Instruct serving as a control. The mechanistic behavioral protocol comprised one deterministic forward pass (selecting the argmax over the A/B token logits) alongside nine stochastic passes sampled at temperature 0.7 across the identical 225-item instrument. While mechanistic coefficients do not directly map onto the 20-model panel results, they provide convergent validity on the underlying test items. Probing value directions from the residual stream on scenario-based items aligns with established literature (Section~\ref{sec:1-2}). However, cross-model representational alignment---specifically whether distinct architectures utilize shared value directions---remains unresolved; the cross-model activation patching and representational comparisons below address this question on the subset of disagreement items.

Vignette selection for the activation dataset was stratified by inter-model disagreement. The dataset was constructed by extracting the 15 items exhibiting the highest inter-model variance within each of the eight (axis \ensuremath{\times} modality) cells, yielding 120 signal items, supplemented by 40 control items characterized by unanimous consensus choices across models (160 items total). The top-15 selection rule was enforced without applying a minimum disagreement threshold; the empirical implications for dataset composition are detailed in Section~\ref{sec:3-11-1}. Internal activation extraction was executed exclusively on the 120-item signal set.

Classifiers were trained on 60 items per cell: the 120-item signal dataset bisects into 60 emancipative and 60 sacred--secular items, with each probe trained exclusively on its corresponding axis. Preprocessing comprised standard scaling followed by principal component analysis (PCA) reducing feature space to 128 components, evaluated via 5-fold stratified cross-validation. The optimal layer for each model-axis pair was identified by maximizing cross-validated accuracy across layers on the training folds; reported accuracies reflect 5-fold cross-validated means without nested cross-validation. Consequently, two-proportion hypothesis tests against majority-class baselines are not reported for specific sample sizes \emph{n}; performance margins over majority-class baselines are reported throughout.

For each model and axis, the difference vector between the mean activations of the two response classes was defined as a candidate value direction. Directional alignment across models was evaluated via cosine similarity: directly within the 4,096-dimensional hidden-state architecture group, and via PCA dimensionality reduction (to 59 components) followed by Procrustes alignment across distinct architectural groups. As a theoretical reference null, two independent uniform unit vectors in \(\mathbb{R}^{d}\) exhibit an expected mean cosine similarity of 0 with a standard deviation of \(1/\sqrt{d}\), corresponding to 0.0156 at d = 4,096.

Representational similarity analysis (RSA) and centered kernel alignment (CKA) were computed layer by layer across all model pairs over the item set. Reported metrics include the mean similarity, maximum similarity, and layer index corresponding to peak CKA alignment.

Causal patching interventions were conducted by injecting a value direction extracted from a source model into the residual stream of a target forward pass at a specified layer, measuring the proportion of items for which the emitted choice reversed. Within-model patching is reported at the layer yielding peak causal effect. Cross-model patching transfers directional vectors between distinct models within the same dimension group, evaluated exclusively on items where source and target models exhibit baseline disagreement (comprising 8 items on the emancipative axis and 9 items on the sacred--secular axis; individual item flips thus shift the response rate by approximately 11--12 percentage points). Cross-model patching across differing architecture dimensions is geometrically invalid and omitted. Asymmetry in causal transfer was evaluated using two-sided Fisher's exact tests, with proportions bounded by 95\% Wilson score confidence intervals.

\FloatBarrier
\section{Results}\label{sec:3}

\subsection{Response yield, run-level agreement and effective sample size}\label{sec:3-1}

Refusals and unparseable outputs, evaluated across raw response cells prior to aggregation, are distributed as follows: English, 2,658 non-A/B cells out of 148,500 (1.79 \%); Russian, 1,539 out of 74,250 (2.07 \%); and the second temperature condition, 1,992 out of 78,750 (2.53 \%). Explicit refusal tokens are rare, occurring in 51 instances in English and zero in the remaining conditions; missingness predominantly comprises blank or unparseable text. Thirty-eight English vignettes yielded zero usable runs, distributed among Alice AI LLM (11), YandexGPT (11), YandexGPT-Lite (11), and gpt-oss-120b (5). This results in four models having 214, 214, 214, and 220 usable vignettes, respectively, while all remaining models retain 225 usable vignettes. This decomposition reproduces the pairwise-comparison denominators derived independently in the permutation-testing pipeline.

Run-level agreement is near-degenerate at temperature 0. Per-model self-consistency ranges from 91.7 \% to 100 \%, with a median near 99.5 \%. Six systems exhibit complete consistency across all evaluated vignettes; the least consistent system, Sonar Pro, demonstrates self-consistency on 91.7 \% of runs. A one-way random-effects intra-cluster correlation on binary run outcomes---treating the vignette as the cluster and excluding refusals from cluster size---spans 0.748 to 1.000 (median 0.985). Corresponding design effects range between 22.7 and 50.0, yielding effective sample sizes between 214 and 296 per model (median 227), compared to 6,338 to 11,200 raw responses per model. The lower boundary of this range corresponds to the three systems with ten unparseable vignettes; no model's internal inconsistency falls into this lower bound. Structuring the analysis around 6,750 independent observations with clustering corrections would artificially inflate the sample size by over an order of magnitude. The dataset effectively comprises approximately 224 vignette-level observations per model, with the 30 or 50 runs per item representing an empirical agreement rate.

Two observations regarding panel composition follow from the response data. Alice AI LLM and YandexGPT produce identical per-vignette choice proportions across all 214 vignettes with valid outputs, with identical raw label multisets on every item; the panel therefore effectively comprises 19 distinct architectures under 20 model designations. Ten design cells are unparseable for both models due to refusal or empty output and were subsequently excluded from joint statistical evaluations.

\subsection{Concordance under the four perturbations}\label{sec:3-2}

\begin{figure}[htbp]
\centering
\includegraphics[width=.96\linewidth,height=.77\textheight,keepaspectratio]{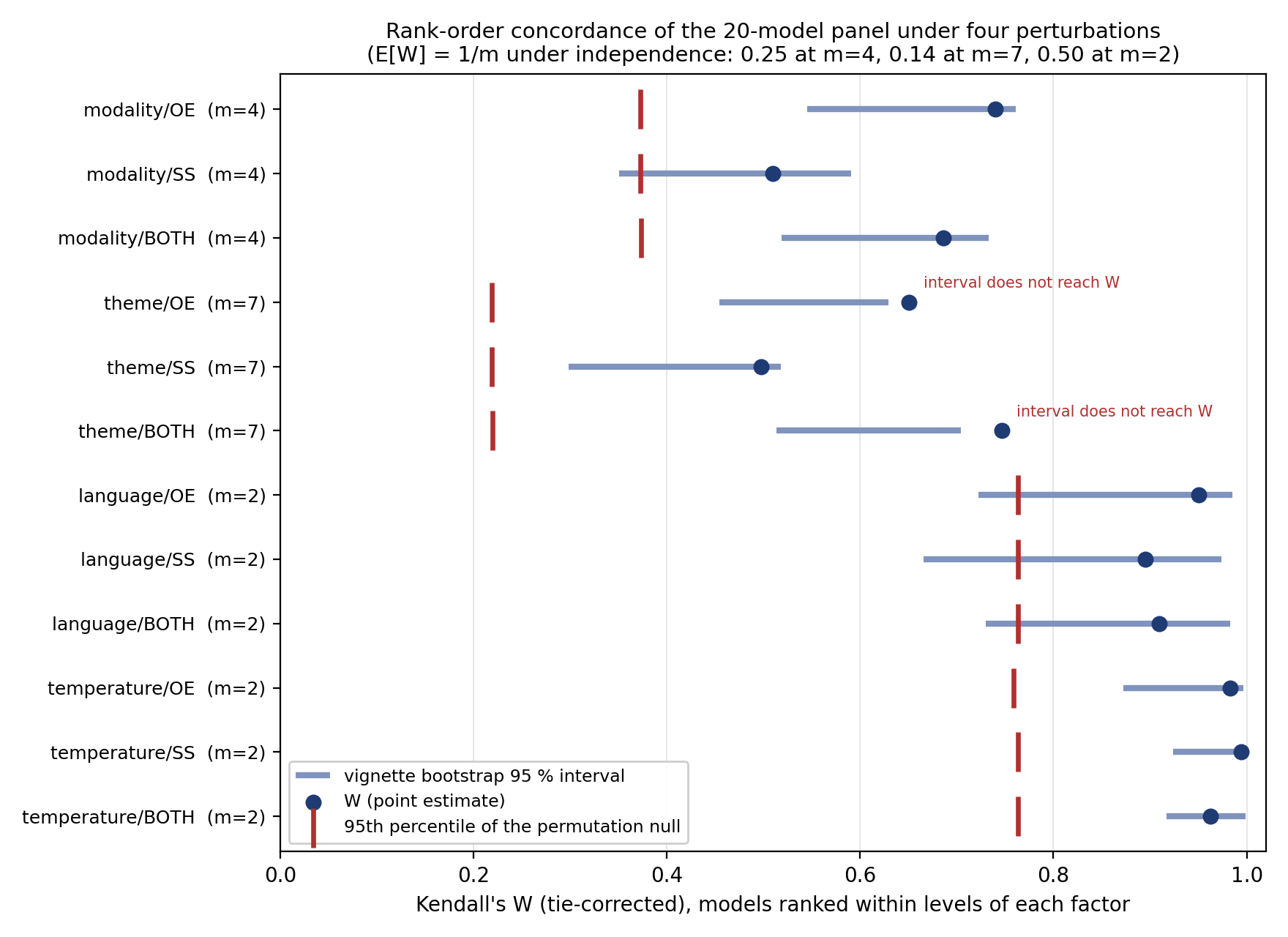}
\caption{Rank-order concordance of the twenty-model panel under the four perturbations, with the empirical null. Each row is one (factor, axis) cell. The blue point is the tie-corrected Kendall's W; the blue bar is the 95 \% percentile interval from 2,000 bootstrap replicates resampling vignettes; the red tick is the 95th percentile of a 20,000-replicate permutation null of independent random rankings, whose mean is 1/m. Two cells are annotated because their percentile interval does not reach their own point estimate, a property of the percentile bootstrap discussed in Section~\ref{sec:3-9}.}
\label{fig:1}
\end{figure}

Every experimental cell significantly exceeds its empirical null distribution: model orderings induced by the benchmark exhibit rank-order concordance above chance expectation across all four perturbation factors, evaluated on both value axes independently as well as pooled. Variations in assigned decision role, thematic scenario domain, interface language, and sampling temperature do not reorder the model panel into substantially divergent performance tiers. However, exceeding a random null represents a baseline criterion; Section~\ref{sec:stability-ceilings} additionally evaluates each factor against the theoretical concordance ceiling attainable under perfect rank stability for this experimental design, which is bounded below 1 for item factors and equals 1 for paired factors.

The two-level factors operate on a different mathematical scale relative to multi-level factors: language and temperature are evaluated at \emph{m} = 2, where chance concordance is 0.50. High concordance coefficients for these factors indicate strong alignment between two rankings rather than double the stability observed for decision modality. Cross-factor comparison requires referencing the null distribution columns in Table~\ref{tab:concordance} (Appendix~\ref{app:b}) and the stability ceilings detailed in Section~\ref{sec:stability-ceilings}.

\begin{figure}[htbp]
\centering
\includegraphics[width=1\linewidth,height=.77\textheight,keepaspectratio]{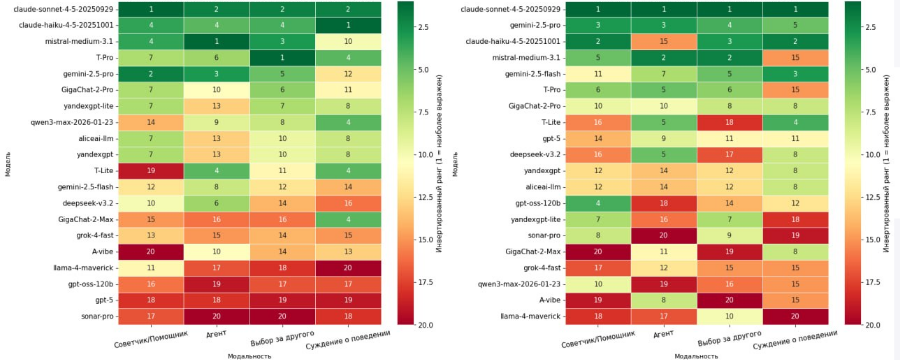}
\caption{Rank of each model within each decision modality in the English condition, on the obedient--emancipative axis (left) and the sacred--secular axis (right). Rank 1 is the highest share of the counted pole (emancipative, secular) and rank 20 the lowest; the colour scale runs from green at rank 1 to red at rank 20. Columns, left to right: Advisor, Agent, Choice on behalf of the person, Judgment of another's behaviour. The column labels and the colour-bar label (``inverted rank, 1 = most pronounced'') are in Russian in the source graphic.}
\label{fig:2}
\end{figure}

Model rank positions exhibit stability at the distribution tails while demonstrating variance in the middle ranks. Ranking models within each decision modality, Claude Sonnet 4.5 occupies the first or second position across all modalities on both axes, whereas Sonar Pro and GPT-5 remain in the bottom quartile on the emancipative axis across all modalities. Conversely, mid-tier models display rank shifts spanning 9 to 14 positions out of 20: T-lite shifts from rank 19 to rank 4 across modalities on the emancipative axis, and gpt-oss-120b shifts from rank 4 to rank 18 on the sacred--secular axis. Figure~\ref{fig:2} illustrates the complete rank matrix across both axes.

\subsection{Level shifts by decision modality and theme}\label{sec:3-3}

\begin{figure}[htbp]
\centering
\includegraphics[width=.82\linewidth,height=.77\textheight,keepaspectratio]{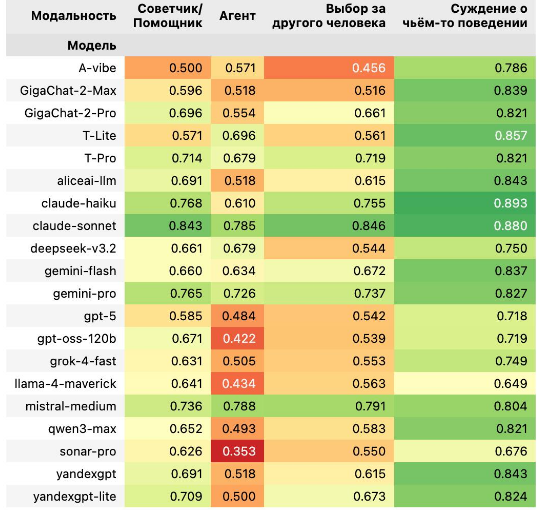}
\caption{Share of responses selecting Option 2, the slot that carries the counted pole in 91.5 \% of polarity-identifiable items, per model and decision modality, English condition, pooled over both axes; green marks higher shares and red lower ones. Column means agree with the pooled modality levels of Section~\ref{sec:3-3} to within 0.006. The column headers are in Russian in the source graphic: Advisor, Agent, Choice on behalf of the person, Judgment of another's behaviour.}
\label{fig:3}
\end{figure}

\begin{figure}[htbp]
\centering
\includegraphics[width=.88\linewidth,height=.77\textheight,keepaspectratio]{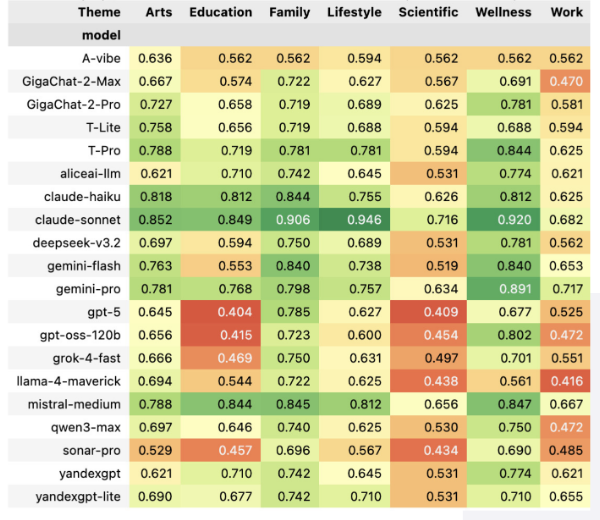}
\caption{Share of responses selecting Option 2, per model and theme, English condition, pooled over both axes; green marks higher shares and red lower ones. Column means agree with the pooled theme levels of Section~\ref{sec:3-3} to within 0.002.}
\label{fig:4}
\end{figure}

While concordance evaluates rank-order agreement, absolute score levels exhibit substantial shifts across experimental conditions. Pooled across the model panel, the mean counted-pole share by decision modality is Advisor 0.669, Agent 0.573, Choice on behalf 0.619, and Judgment of another's behaviour 0.797, reflecting an absolute range of 0.224 on a [0, 1] scale. Across thematic domains, pooled scores range from Family (0.756) and Wellness (0.753) to Scientific (0.549) and Work (0.578), yielding a domain spread of 0.207. The direction of the modality effect indicates that models express significantly higher emancipative tendencies when judging external behavior (0.797) compared to acting agentically on behalf of a user (0.573). Figures~\ref{fig:3} and \ref{fig:4} display individual model response proportions across decision modalities and thematic domains, respectively.

\subsection{Role sensitivity per model}\label{sec:3-4}

Section~\ref{sec:3-3} demonstrates that absolute response levels vary by decision role, whereas Section~\ref{sec:3-2} establishes overall rank-order concordance. These findings indicate heterogeneous model sensitivity to assigned roles: because decision role is held uniform across the panel, non-uniform score shifts across models are required to alter relative rankings. Table~\ref{tab:role-sensitivity} quantifies this inter-model variability in role sensitivity.

\begin{table}[htbp]
\centering
\small
\renewcommand{\arraystretch}{1.12}
\caption{Range of the mean counted-pole share across the four decision modalities, per model and axis, English condition, on a scale bounded by zero and one. Computed from the per-vignette proportions with refusals dropped and the duplicated row removed.}\label{tab:role-sensitivity}
\begin{tabular}{@{}lrr@{}}
\toprule
Model & OE range & SS range \\
\midrule
DeepSeek v3.2 & 0.071 & 0.356 \\
Mistral Medium 3.1 & 0.073 & 0.094 \\
Claude Sonnet 4.5 & 0.110 & 0.105 \\
Gemini 2.5 Pro & 0.119 & 0.114 \\
Gemini 2.5 Flash & 0.244 & 0.213 \\
T-Pro & 0.250 & 0.179 \\
Claude Haiku 4.5 & 0.250 & 0.315 \\
GPT-5 & 0.311 & 0.289 \\
Grok 4 Fast & 0.314 & 0.303 \\
gpt-oss-120b & 0.346 & 0.290 \\
Llama 4 Maverick & 0.352 & 0.172 \\
GigaChat 2 Pro & 0.357 & 0.229 \\
T-Lite & 0.357 & 0.393 \\
Qwen3 Max & 0.370 & 0.315 \\
Sonar Pro & 0.389 & 0.257 \\
A-vibe & 0.393 & 0.500 \\
Alice AI LLM & 0.426 & 0.292 \\
YandexGPT Lite & 0.426 & 0.208 \\
YandexGPT & 0.426 & 0.292 \\
GigaChat 2 Max & 0.461 & 0.393 \\
\bottomrule
\end{tabular}
\end{table}

On the emancipative axis, role sensitivity ranges across models from 0.071 to 0.461 (a 6.5-fold difference); on the sacred--secular axis, it ranges from 0.094 to 0.500 (a 5.3-fold difference), with median ranges of 0.349 and 0.289, respectively. This variation spans a continuous distribution: four models exhibit ranges below 0.12 on the emancipative axis, while five models exceed 0.39. Three architectures demonstrate high role invariance across both axes (Mistral Medium 3.1, Claude Sonnet 4.5, and Gemini 2.5 Pro, all \ensuremath{\leq} 0.12), whereas three architectures exhibit high role sensitivity across both axes (GigaChat 2 Max, A-vibe, and T-Lite, all \ensuremath{\geq} 0.35). Remaining models display axis-asymmetric sensitivity: DeepSeek v3.2 exhibits the highest stability on the emancipative axis (0.071) alongside moderate variation on the sacred--secular axis (0.356). A-vibe records an absolute range of 0.500 on the sacred--secular axis---representing a shift of half the measurement scale between its extreme roles. Scores for Alice AI LLM and YandexGPT are identical, reflecting the output equivalence noted in Section~\ref{sec:3-1}.

\subsection{Stability relative to empirical ceilings}\label{sec:stability-ceilings}

For paired experimental factors, the theoretical stability ceiling equals 1.000: the median rank correlation between split-half runs of a single evaluation condition is 1.000 on both axes due to the near-deterministic output at temperature 0. For item-level factors, the theoretical ceiling is bounded below 1.000, as formalised in Section~\ref{sec:2-8}. Table~\ref{tab:ceilings} compares observed mean pairwise rank correlations against their theoretical stability bounds.

\begin{table}[htbp]
\centering
\small
\renewcommand{\arraystretch}{1.12}
\caption{Observed mean pairwise rank correlation between the levels of each factor against the ceiling obtained under perfect stability. \(\bar{\rho}\) is the mean pairwise rank correlation between the levels of the factor, the last column of Table~\ref{tab:concordance}, related to the reported coefficient by W = ((m\ensuremath{-}1)\(\bar{\rho}\) + 1)/m. Theme and modality are on the twenty-model panel; language and temperature on the eleven models administered in both conditions. Ceilings from 800 rotations for the item factors and 600 run splits for the paired factors.}\label{tab:ceilings}
\begin{tabular}{@{}llrrrr@{}}
\toprule
Factor & Axis & \makecell{\(\bar{\rho}\)\\observed} & \makecell{\(\bar{\rho}\)\\ceiling} & \makecell{Share of\\ceiling} & \makecell{Percentile in\\ceiling distribution} \\
\midrule
Temperature & SS & 0.989 & 1.000 & 0.989 & below the 1st \\
Temperature & OE & 0.966 & 1.000 & 0.966 & below the 1st \\
Theme & SS & 0.413 & 0.436 & 0.947 & 29.2 \\
Language & OE & 0.901 & 1.000 & 0.901 & below the 1st \\
Theme & OE & 0.592 & 0.685 & 0.864 & 1.2 \\
Modality & OE & 0.653 & 0.798 & 0.818 & 0.1 \\
Language & SS & 0.791 & 1.000 & 0.791 & below the 1st \\
Modality & SS & 0.347 & 0.559 & 0.620 & 0.0 \\
\bottomrule
\end{tabular}
\end{table}

All evaluated factors induce measurable rank adjustments, with decision role exerting the largest impact on ranking stability. Seven of eight factor-by-axis cells fall below the 1.5th percentile of their empirical stability ceilings; the sole exception is thematic domain on the sacred--secular axis (29th percentile), which is statistically consistent with complete rank stability. Factor sensitivity follows a hierarchical gradient from temperature (highest stability), followed by language and thematic domain, to decision role (lowest stability).

Restricting statistical evaluation to the eleven models administered across all four conditions reduces relative concordance values across all factors, as this subset is concentrated with closely ranked models that are more susceptible to rank reordering. Specifically, thematic concordance decreases to 0.824 (OE) and 0.736 (SS), while modality concordance decreases to 0.627 (OE) and 0.570 (SS). However, the relative hierarchy among factors remains invariant: decision role preserves less theoretical rank stability than thematic domain across both axes and panel subsets, whereas sampling temperature preserves the highest proportion of stability. Intermediate factor rankings exhibit minor panel dependence: thematic domain outranks language on the 20-model panel for the sacred--secular axis, but not on the 11-model restricted panel.

\subsection{Language of administration}\label{sec:3-6}

In the paired cross-lingual evaluation across eleven models using identical translated items, language substitution induces systematic score shifts without substantially reordering model rankings. Administration in Russian shifts ten of eleven models toward the counted pole on the emancipative axis and seven of eleven on the sacred--secular axis, with mean absolute score displacements of 0.029 and 0.036, respectively. The maximum observed displacement occurs in DeepSeek v3.2 (+0.071 on the sacred--secular axis). Within-condition split-run sampling noise yields a median displacement of 0.000 (maximum 0.0013), confirming that cross-lingual shifts significantly exceed baseline sampling variance.

The statistical significance of cross-lingual shifts depends on the designated unit of analysis. Aggregating at the vignette level (Section~\ref{sec:2-4}), two of twenty-two (model \ensuremath{\times} axis) contrasts reach uncorrected significance via paired Wilcoxon signed-rank tests across 112 vignettes: Qwen3 Max on the sacred--secular axis (\emph{p} = 0.007) and GigaChat 2 Max on the emancipative axis (\emph{p} = 0.043); neither remains significant under Bonferroni correction across the twenty-two tests. Evaluated at the individual-response level (\textasciitilde{}3,200 observations per model per axis), twelve contrasts reach uncorrected significance and eleven survive Bonferroni correction. These metrics address distinct inferential units; item-generalizable benchmark evaluations require treating the vignette as the primary sampling unit, and cross-lingual shifts are accordingly reported as absolute score displacements alongside baseline noise bounds.

Rank order is highly preserved across language conditions, achieving a pooled concordance of \emph{W} = 0.909 and a mean pairwise Spearman correlation of 0.818 between English and Russian league tables (OE: W = 0.951, \(\bar{\rho}\) = 0.901; SS: W = 0.895, \(\bar{\rho}\) = 0.791). Relative to the paired stability ceiling of 1.000 (Section~\ref{sec:stability-ceilings}), the cross-lingual ordering achieves 0.901 of theoretical stability on the emancipative axis and 0.791 on the sacred--secular axis, with both values falling below the 1st percentile of the permutation ceiling. Thus, while interface language induces minor rank adjustments, the underlying value profile remains substantially consistent across languages.

Observed shifts do not systematically align with developer geographic origin. Both Yandex-developed models exhibit small cross-lingual score differences (+0.018 and +0.009 on OE; \ensuremath{-}0.010 and \ensuremath{-}0.020 on SS), none of which reach significance at either aggregation level. Conversely, the four largest language-induced displacements occur in DeepSeek v3.2, GigaChat 2 Max, Qwen3 Max, and GigaChat 2 Pro---a set spanning both Russian and non-Russian development organizations. Cross-lingual displacement magnitude is thus uncorrected with developer origin.

\subsection{Sampling temperature}\label{sec:3-7}

Comparing temperature 0 against vendor-recommended sampling temperatures yields no statistically significant differences for any of the eleven tested models across either axis. Sampling temperature demonstrates the highest rank stability among all evaluated factors: pooled \emph{W} = 0.962, with a mean pairwise Spearman correlation of 0.925 (OE: W = 0.983, \(\bar{\rho}\) = 0.966; SS: W = 0.994, \(\bar{\rho}\) = 0.989). Relative to the paired stability ceiling of 1.000, temperature variation achieves 0.966 of theoretical stability on the emancipative axis and 0.989 on the sacred--secular axis. Consequently, temperature setting represents a highly benign factor with negligible impact on relative model evaluation.

\subsection{Reasoning mode}\label{sec:3-8}

\begin{figure}[htbp]
\centering
\includegraphics[width=.98\linewidth,height=.77\textheight,keepaspectratio]{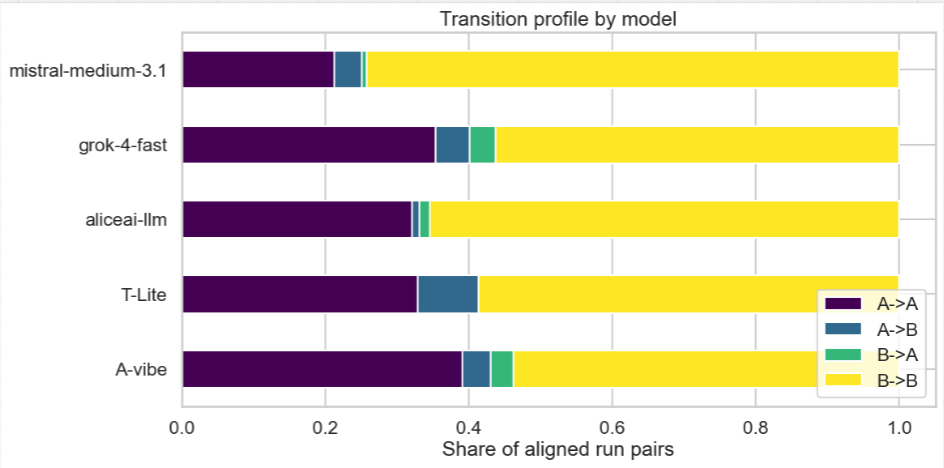}
\caption{Transition profile between the base-mode answer and the reasoning-mode answer for the five models of the reasoning condition: the share of aligned run pairs in which the answer stayed at A, stayed at B, moved from A to B, or moved from B to A.}
\label{fig:5}
\end{figure}

Final choice selections in reasoning mode align with base-mode selections in 90 \% to 97 \% of items across models, peaking at 97 \% for Alice AI LLM. Figure~\ref{fig:5} illustrates choice transition profiles across models (persistent choice A, persistent choice B, or choice switching). Pairwise cosine similarity between mean reasoning-trace embedding centroids ranges from 0.576 to 0.733, indicating shared semantic structure across model reasoning traces without exact equivalence. A gradient-boosted classifier trained on reasoning-trace embeddings predicts final value selections with 0.854 accuracy, 0.812 balanced accuracy, and 0.885 ROC-AUC, confirming that generated reasoning traces contain strong predictive signal regarding final choice selection.

\subsection{Sensitivity to analysis specification, and two properties of the intervals}\label{sec:3-9}

An alternative data pipeline imputing non-A/B response cells with the row modal response affect 1,518 English cells (1.02 \%) and shifts Kendall's \emph{W} by \ensuremath{\leq} 0.0051 across all cells. Across eight analytical specifications per cell, concordance metrics remain stable: modality/OE ranges from 0.7161 to 0.7406; modality/SS from 0.5040 to 0.5139; pooled modality from 0.6684 to 0.6858; theme/OE from 0.6305 to 0.6508; theme/SS from 0.4970 to 0.4976; and pooled theme from 0.7310 to 0.7467. Evaluating all 48 specification permutations yields consistent bounds, demonstrating that primary concordance conclusions are robust to analytical specification choices.

Interpretation of the bootstrap percentile intervals in Table~\ref{tab:concordance} requires noting two structural properties. Because vignette resampling introduces random noise into level means, bootstrap resampling systematically attenuates concordance, causing bootstrap means to fall below empirical point estimates (e.g., modality/OE point estimate 0.740 vs. bootstrap mean 0.664). This suggests that the lower interval bound provides an informative lower limit on stability, whereas the upper bound should not be interpreted as an upper limit on W. Additionally, two of twelve percentile intervals exclude their empirical point estimates: theme/OE (W = 0.6508, CI [0.4546, 0.6294]) and pooled theme (W = 0.7467, CI [0.5135, 0.7045]). We report these raw percentile intervals directly; conversely, pooled temperature yields a bootstrap mean (0.9746) above its point estimate (0.9624), ruling out unidirectional attenuation.

\subsection{Internal consistency}\label{sec:3-10}

Concordance quantifies rank-order preservation across experimental conditions rather than internal test reliability within a single condition.

\subsection{Mechanistic sub-study}\label{sec:3-11}

\subsubsection{Behavioural positions and composition of the activation set}\label{sec:3-11-1}

Under the sub-study protocol, model response proportions at the counted pole are: gemma (86.2 \% OE, 70.2 \% SS), T-lite (73.3 \% OE, 63.6 \% SS), a-vibe (66.1 \% OE, 57.2 \% SS), and Qwen2.5 control (51.3 \% OE, 37.2 \% SS). The four primary models exhibit choice agreement on 61.8 \% of deterministic evaluations, with a mean inter-model variance of 0.0844.

\begin{figure}[htbp]
\centering
\includegraphics[width=1\linewidth,height=.77\textheight,keepaspectratio]{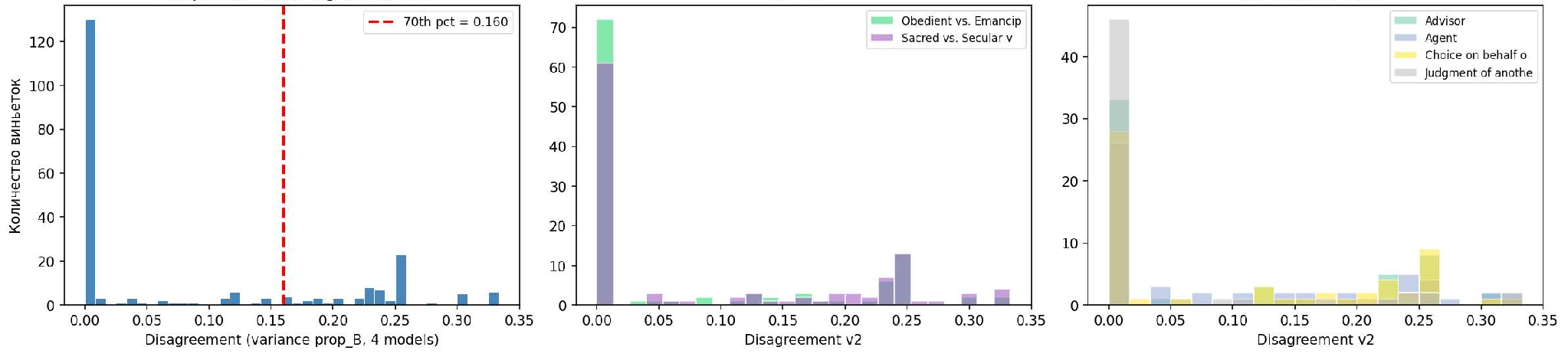}
\caption{Between-model disagreement across the 225 items in the mechanistic sub-study, measured as the variance of the Option 2 share across the four main models: the overall distribution, with its 70th percentile (0.160) marked by the dashed line (left), and the same distribution split by axis (centre) and by decision modality (right). The vertical-axis label of the left panel (``number of vignettes'') is in Russian in the source graphic.}
\label{fig:6}
\end{figure}

Between-model disagreement occurs on 106 of the 225 items (47 \%), yielding a median variance of 0.0000 (Figure~\ref{fig:6}). Because top-15 item selection was executed without enforcing a minimum variance threshold, 25 of the 120 signal items exhibit zero inter-model disagreement, leaving 95 variance-bearing signal items (79 \%). Internal activations were extracted exclusively for the 120 signal items; the 40 control items were excluded from activation extraction, and no specificity evaluation was performed on control items.

\subsubsection{Linear probes}\label{sec:3-11-2}

\begin{figure}[htbp]
\centering
\includegraphics[width=1\linewidth,height=.77\textheight,keepaspectratio]{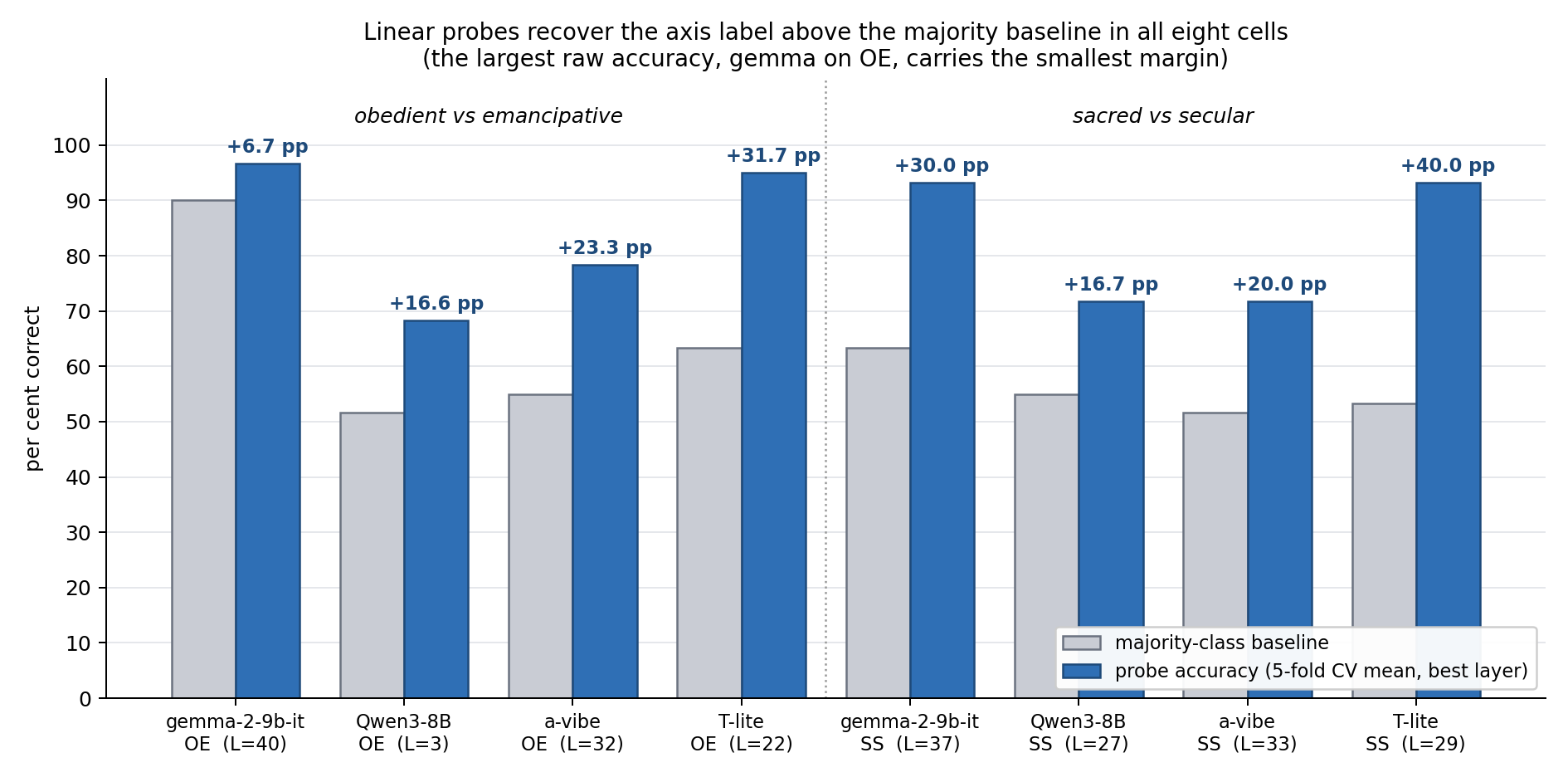}
\caption{Linear probes recover the axis label above the majority-class baseline in all eight model \ensuremath{\times} axis cells; the margin is the quantity to read. Grey bars are the majority-class share of the 60 items in that cell; blue bars are the 5-fold cross-validated probe accuracy at the layer of peak accuracy, with the layer index given under each label; the annotation is the margin in percentage points.}
\label{fig:7}
\end{figure}

\begin{table}[htbp]
\centering
\footnotesize
\setlength{\tabcolsep}{3.5pt}
\caption{Probe accuracy at the layer of peak cross-validated accuracy against the majority-class baseline, per model and axis, with the margin in percentage points (pp).}\label{tab:probes}
\begin{tabular}{@{}lrrrrrrrr@{}}
\toprule
& \multicolumn{4}{c}{Obedient--emancipative} & \multicolumn{4}{c}{Sacred--secular} \\
\cmidrule(lr){2-5}\cmidrule(l){6-9}
Model & Layer & Accuracy & Baseline & Margin & Layer & Accuracy & Baseline & Margin \\
\midrule
gemma-2-9b-it & 40 & 96.7 \% & 90.0 \% & +6.7 pp & 37 & 93.3 \% & 63.3 \% & +30.0 pp \\
Qwen3-8B & 3 & 68.3 \% & 51.7 \% & +16.6 pp & 27 & 71.7 \% & 55.0 \% & +16.7 pp \\
a-vibe & 32 & 78.3 \% & 55.0 \% & +23.3 pp & 33 & 71.7 \% & 51.7 \% & +20.0 pp \\
T-lite & 22 & 95.0 \% & 63.3 \% & +31.7 pp & 29 & 93.3 \% & 53.3 \% & +40.0 pp \\
\bottomrule
\end{tabular}
\end{table}

\begin{figure}[p]
\centering
\includegraphics[width=1\linewidth,height=.77\textheight,keepaspectratio]{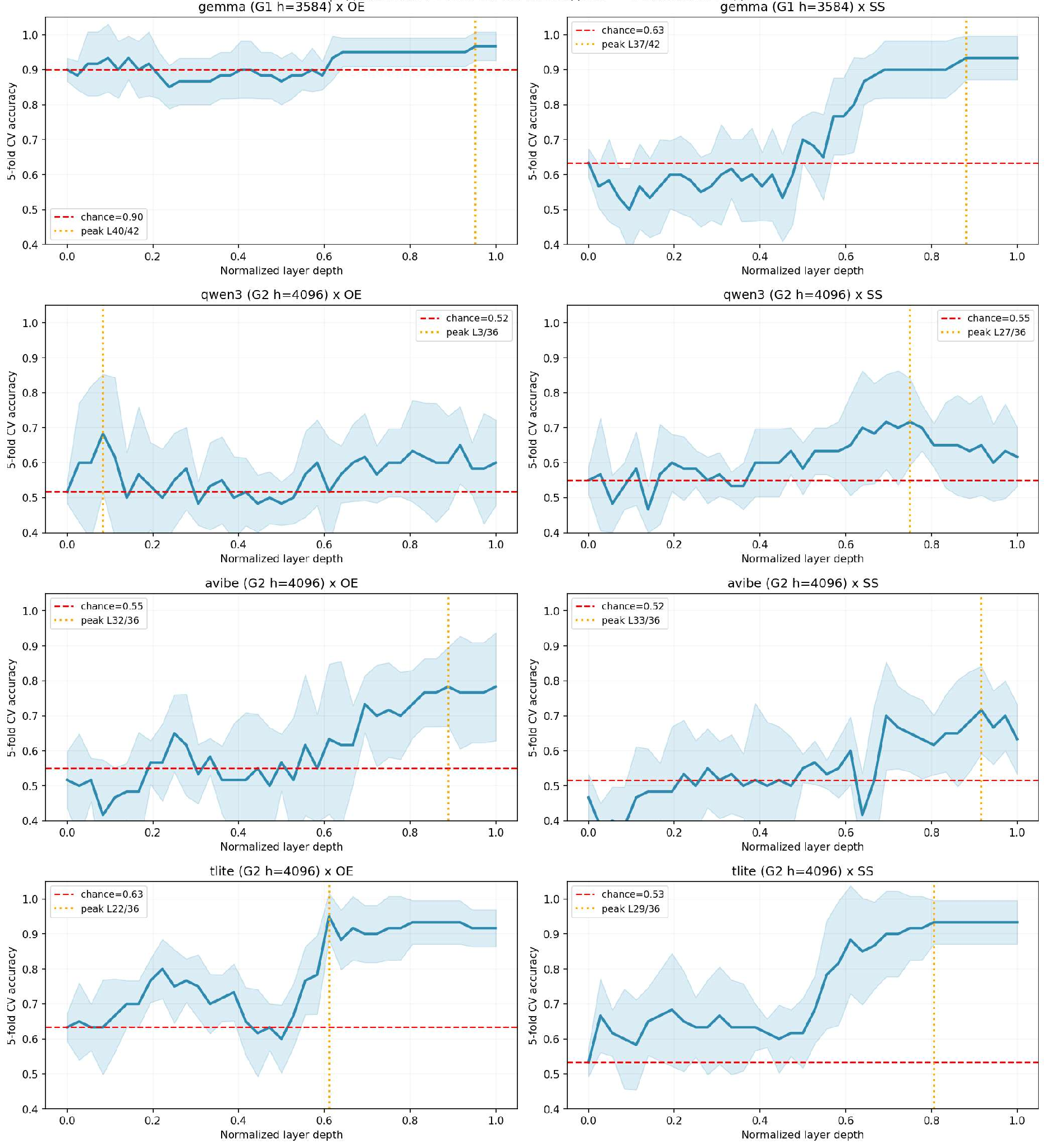}
\caption{Five-fold cross-validated probe accuracy by normalised layer depth for the four main models of the mechanistic sub-study (rows), on the obedient--emancipative axis (left column) and the sacred--secular axis (right column). The dashed red line is the majority-class baseline of the cell and the dotted orange line the layer of peak accuracy reported in Table~\ref{tab:probes}.}
\label{fig:8}
\end{figure}

Linear probes trained on residual stream activations recover axis labels above majority-class baselines across all eight (model \ensuremath{\times} axis) cells, achieving accuracies between 68.3 \% and 96.7 \% against baselines of 51.7 \% to 90.0 \%. These results demonstrate that value axis distinctions are linearly decodable from internal model states across all evaluated open-weight architectures. Probe accuracy must be evaluated relative to majority-class baselines: gemma achieves 96.7 \% accuracy on the emancipative axis against a 90.0 \% baseline (a +6.7 percentage point margin), reflecting gemma's high behavioral selection rate (\textasciitilde{}86 \%) for emancipative choices; conversely, T-lite achieves 93.3 \% accuracy on the sacred--secular axis against a 53.3 \% baseline (+40.0 percentage point margin). Figure~\ref{fig:7} displays cross-validated classification margins across models and axes, and Figure~\ref{fig:8} details accuracy profiles across normalized layer depth.

\subsubsection{Difference-of-means directions}\label{sec:3-11-3}

Direct cosine similarities between difference-of-means vectors within the 4,096-dimensional architecture group are: Qwen3--a-vibe (0.003 OE, \ensuremath{-}0.021 SS); Qwen3--T-lite (0.009 OE, 0.006 SS); and a-vibe--T-lite (0.025 OE, 0.296 SS). Across architecture groups (following 59-component PCA and Procrustes alignment), direction similarities measure: gemma--Qwen3 (0.555 OE, \ensuremath{-}0.426 SS); gemma--a-vibe (0.488 OE, 0.436 SS); and gemma--T-lite (0.782 OE, 0.524 SS).

Four of six within-group directional similarities are statistically indistinguishable from orthogonal random vectors. Evaluated relative to the isotropic null standard deviation (\ensuremath{\sigma} \ensuremath{\approx} 0.0156 at d = 4,096; Section~\ref{sec:2-10}), within-group alignments correspond to z-scores of: Qwen3--a-vibe (0.2 OE, \ensuremath{-}1.3 SS); Qwen3--T-lite (0.6 OE, 0.4 SS); and a-vibe--T-lite (1.6 OE, 18.9 SS). The single strong alignment occurs between the two Russian fine-tuned models on the sacred--secular axis (z = 18.9), which cannot be isolated from base-model inheritance.

Cross-group similarity metrics exhibit higher nominal values but carry structural qualifications: dimensionality reduction to 59 principal components combined with Procrustes alignment can induce alignment artifacts in sample-limited regimes. Evaluating within-group and cross-group results jointly, the hypothesis that internal value directions are more strongly aligned within architectural families than across families is not supported.

\subsubsection{Representational similarity and centred kernel alignment}\label{sec:3-11-4}

\begin{table}[htbp]
\centering
\small
\renewcommand{\arraystretch}{1.12}
\caption{Representational similarity analysis (RSA) and centred kernel alignment (CKA) between model pairs over the item set, with the layer of peak CKA.}\label{tab:representations}
\begin{tabular}{@{}lrrrrr@{}}
\toprule
Pair & RSA mean & RSA max & CKA mean & CKA max & \makecell{Peak CKA\\layer} \\
\midrule
gemma--Qwen3 & 0.630 & 0.899 & 0.701 & 0.964 & 19 \\
gemma--a-vibe & 0.024 & 0.085 & 0.107 & 0.133 & 34 \\
gemma--T-lite & 0.689 & 0.901 & 0.705 & 0.952 & 21 \\
Qwen3--a-vibe & 0.030 & 0.097 & 0.090 & 0.113 & 1 \\
Qwen3--T-lite & 0.693 & 0.958 & 0.730 & 0.979 & 1 \\
a-vibe--T-lite & 0.037 & 0.077 & 0.097 & 0.126 & 29 \\
\bottomrule
\end{tabular}
\end{table}

Representational similarity analysis across the item set partitions the models into distinct clusters: mean RSA spans 0.024 to 0.693, and mean CKA spans 0.090 to 0.730. Model pairs involving a-vibe display similarity metrics an order of magnitude lower than non-a-vibe pairs (a-vibe pairs: RSA 0.024--0.037, CKA 0.090--0.107; non-a-vibe pairs: RSA 0.630--0.693, CKA 0.701--0.730). This divergence tracks specific model architectures: gemma (from a distinct model family) exhibits representational similarity to Qwen3 and T-lite comparable to their mutual similarity.

Within-model causal activation patching at peak layers flips emitted choices at the following rates: gemma (66.7 \% OE at layer 0, 50.0 \% SS at layer 2); Qwen3 (76.0 \% OE at layer 0, 100.0 \% SS at layer 0); a-vibe (88.0 \% OE at layer 0, 96.0 \% SS at layer 0); and T-lite (90.9 \% OE at layer 13, 76.0 \% SS at layer 14). Writing extracted value directions into residual streams reverses choice outputs at rates between 50 \% and 100 \% across all models and axes, establishing a causal relationship between decodable activation directions and behavioral choice emission within individual models.

\begin{figure}[htbp]
\centering
\includegraphics[width=1\linewidth,height=.77\textheight,keepaspectratio]{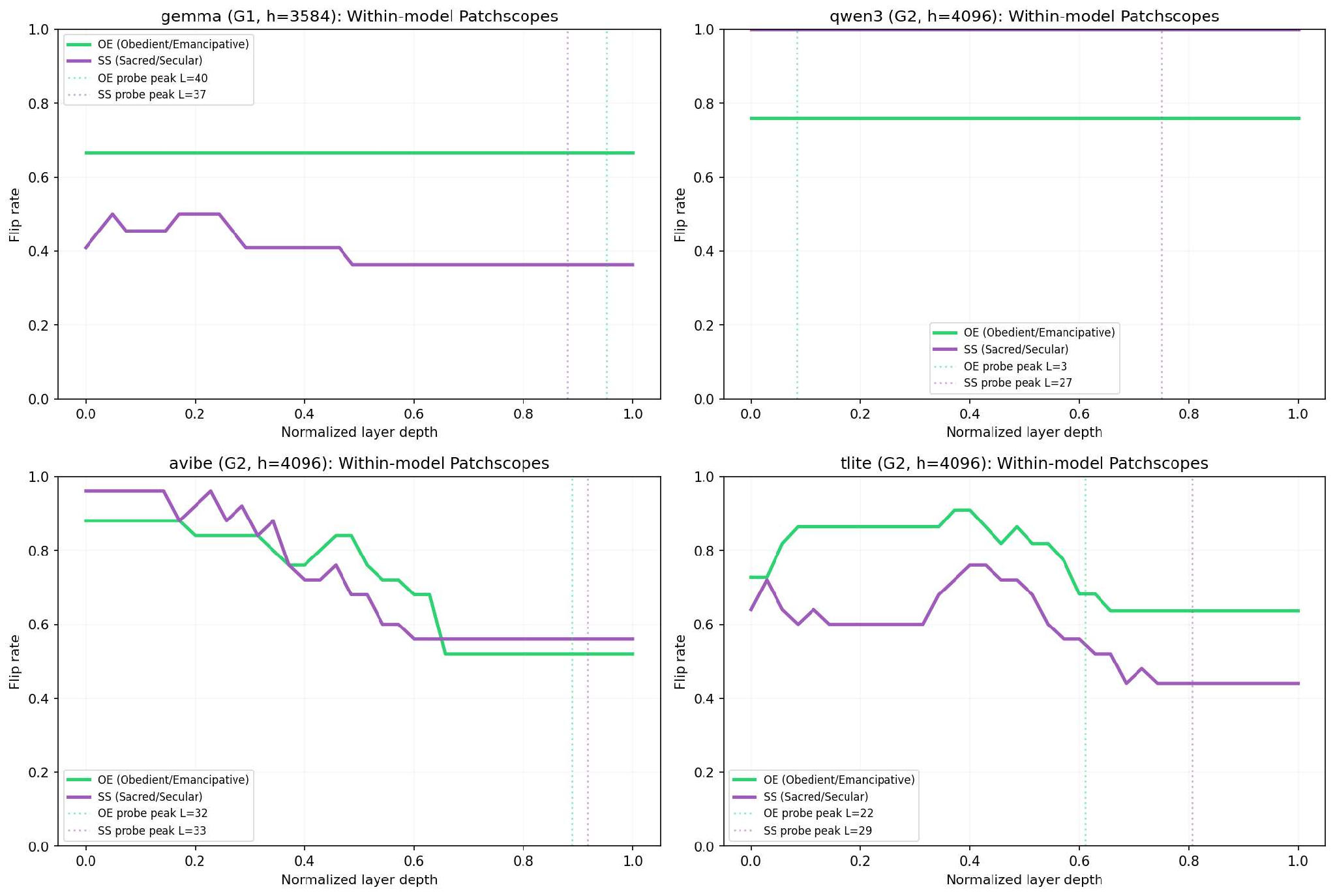}
\caption{Within-model activation patching: the share of items on which the emitted choice flips when the value direction is written into the residual stream, by normalised layer depth, for each of the four main models, on the obedient--emancipative axis (green) and the sacred--secular axis (purple); the dotted vertical lines mark the probe peak layers of Table~\ref{tab:probes}. T-lite peaks in the middle layers; the other three models reach their peak at the embedding layer.}
\label{fig:9}
\end{figure}

Causal depth profiles vary across architectures: T-lite achieves peak intervention effect in intermediate layers, whereas flip rates for the remaining three models remain uniform across network depth, indicating that target value signals are present at the embedding layer (Figure~\ref{fig:9}).

\begin{figure}[htbp]
\centering
\includegraphics[width=1\linewidth,height=.77\textheight,keepaspectratio]{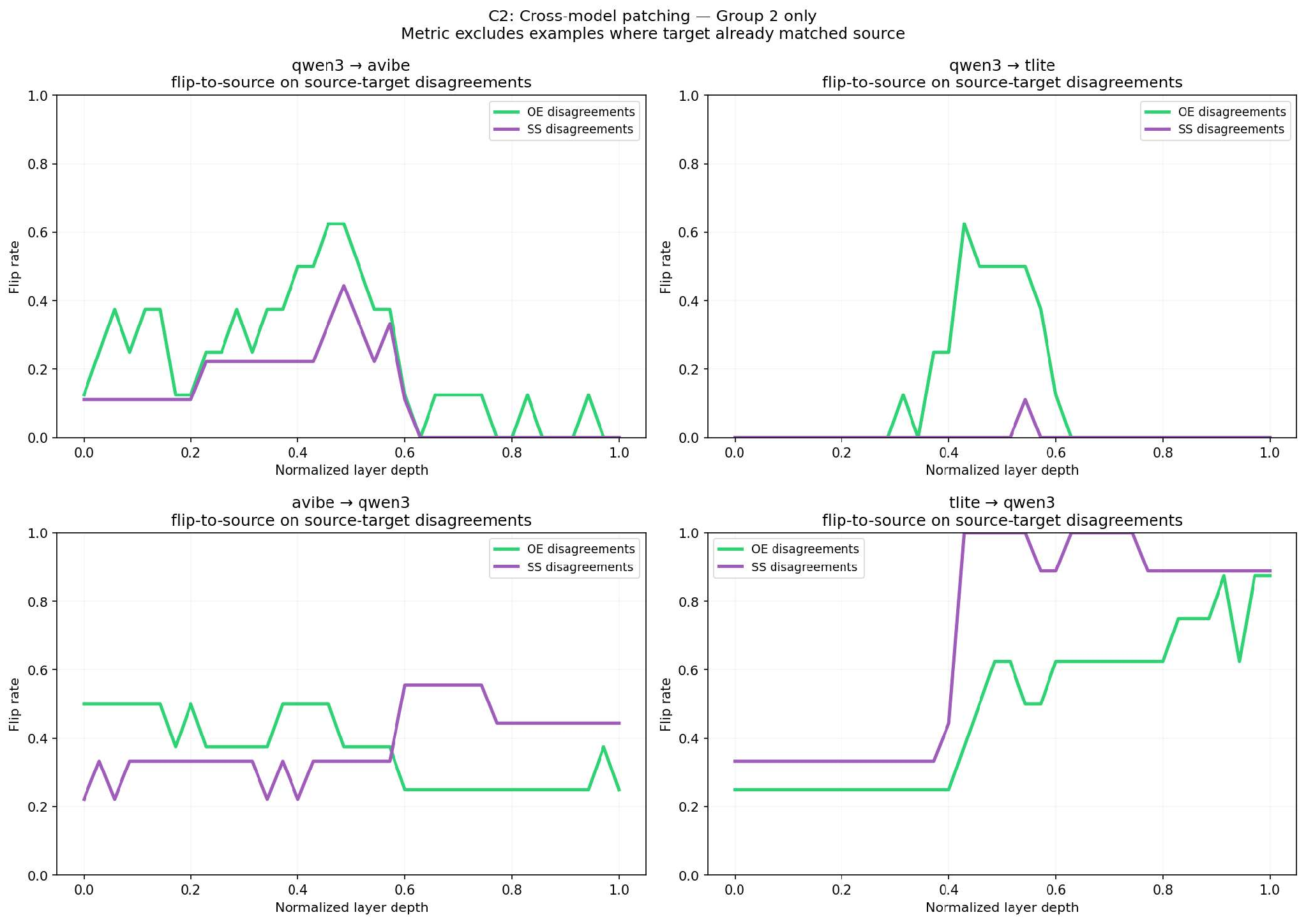}
\caption{Cross-model activation patching within the 4096-dimensional architecture group: the share of source--target disagreement items on which the target's choice flips to the source's choice when the source direction is written into the target, by normalised layer depth, for the four source--target pairs, on the obedient--emancipative axis (green, n = 8 items) and the sacred--secular axis (purple, n = 9).}
\label{fig:10}
\end{figure}

Cross-model activation patching---evaluated strictly on items where source and target models disagree---yields the following choice flip rates. On the emancipative axis (n = 8 items): Qwen3\ensuremath{\rightarrow}a-vibe (62.5 \% at layer 16); Qwen3\ensuremath{\rightarrow}T-lite (62.5 \% at layer 15); a-vibe\ensuremath{\rightarrow}Qwen3 (50.0 \% at layer 0); and T-lite\ensuremath{\rightarrow}Qwen3 (87.5 \% at layer 32). On the sacred--secular axis (n = 9 items): Qwen3\ensuremath{\rightarrow}a-vibe (44.4 \% at layer 17); Qwen3\ensuremath{\rightarrow}T-lite (11.1 \% at layer 19); a-vibe\ensuremath{\rightarrow}Qwen3 (55.6 \% at layer 21); and T-lite\ensuremath{\rightarrow}Qwen3 (100.0 \% at layer 15). Causal vector transfer across models is feasible and exhibits directional asymmetry: vector injection from T-lite into Qwen3 flips choices substantially more frequently than injection from Qwen3 into T-lite. Figure~\ref{fig:10} displays cross-model patching flip rates across normalized layer depth.

Cross-model transfer asymmetry is statistically supported on the sacred--secular axis: a two-sided Fisher's exact test comparing T-lite\ensuremath{\rightarrow}Qwen3 against Qwen3\ensuremath{\rightarrow}T-lite (9/9 vs. 1/9 flips) yields \emph{p} = 4.11 \ensuremath{\times} 10\textsuperscript{-4}. The corresponding emancipative pair (7/8 vs. 5/8 flips) is not significant (\emph{p} = 0.569). Due to small item sample sizes (n = 8--9), 95 \% Wilson score confidence intervals remain wide (9/9: [70 \%, 100 \%]; 1/9: [2 \%, 44 \%]).

\FloatBarrier
\section{Discussion}\label{sec:4}

\subsection{Principal findings and their scope}\label{sec:4-1}

The model ordering induced by this benchmark demonstrates rank-order concordance significantly above the empirical null across all four perturbation factors, both value axes, and eight analytical specifications, regardless of whether refusals are excluded or imputed. This concordance indicates that the elicitation procedure is robust against incidental variations in prompt formulation. However, concordance establishes relative rank-order stability rather than magnitude precision; while model positioning at the distribution extremes remains stable, minor rank-order variations in the middle of the panel should not be interpreted as statistically significant differences.

Performance diverges across the two axes: concordance on the emancipative axis is systematically higher than on the sacred--secular axis across both item factors. Conversely, the mechanistic sub-study demonstrates the opposite asymmetry in representation transfer, which is statistically supported on the sacred--secular axis but not on the emancipative axis. Administration factors behave predictably: language, sampling temperature, and reasoning mode induce minimal variance in model profiles. Notably, pretraining corpus composition does not predict language-induced shifts; models with predominantly Russian pretraining data do not exhibit statistically significant shifts across language conditions.

\subsection{Rank stability and level invariance are distinct claims}\label{sec:4-2}

A key insight from these results is that absolute score levels shift while relative rank ordering is preserved. Consequently, two distinct conclusions emerge: the claim that model rankings survive environmental perturbations is supported, whereas the claim of measurement invariance across conditions is explicitly rejected. A benchmark score cannot be interpreted without specifying the thematic and modality composition of the evaluation set. Reporting a model's emancipative score without contextualizing its item composition omits factors whose extreme levels account for a shift exceeding 20\% of the measurement scale. The full specification of our factorial design (Section~\ref{sec:2-1}) ensures internal comparability across evaluations.

This distinction is critical when comparing our findings with prior work. Trhlik et al. (2026) demonstrate that deployment context exerts a stronger influence on expressed preferences than prompt paraphrasing---a finding consistent with the 0.224 scale spread observed across decision roles in our experiments. While Trhlik et al. caution against treating any single deployment context as a neutral baseline, our analysis distinguishes between absolute score shifts and rank-order preservation. Although absolute scores shift substantially across decision roles, the overall panel ranking remains stable. Evaluating solely absolute level shifts might suggest that the benchmark is unreliable, whereas evaluating solely rank concordance conceals the dependency of scores on role composition. Both perspectives are essential for a complete evaluation.

This distinction clarifies why concordance coefficients in similar numerical ranges yield different interpretations across studies. Trhlik et al. report a mean \emph{W} of 0.66 across eleven traits (spanning 0.36 for fear to 0.90 for surprise) across nine models, concluding that model rankings lack context invariance. Our decision modality concordance metrics fall into a comparable numerical range (0.740 and 0.510 across the two axes; 0.686 pooled), yet support the conclusion that relative rank ordering is preserved. This divergent interpretation stems from three factors: evaluation thresholds, target constructs, and score dispersion across models.

First, Trhlik et al. evaluate concordance against strict measurement invariance, whereas our analysis evaluates concordance against an empirical permutation null calibrated to the factor levels and panel size (0.373 for four modalities across twenty models; 0.219 for seven themes). A concordance coefficient of 0.740 against a chance baseline of 0.250 demonstrates substantial rank preservation relative to chance expectation. Second, while Trhlik et al. perform exploratory classification of free-text outputs, our benchmark evaluates value axes using a structured, purpose-built scenario instrument. Third, the degree of score dispersion across models differs dramatically between the two studies. Trhlik et al. evaluate models that cluster tightly (median absolute difference of 0.3 percentage points, with Big Five ranges never exceeding 1.6 percentage points), meaning that minor output perturbations easily induce rank inversions among virtually indistinguishable models. In contrast, our panel exhibits substantial score dispersion (counted-pole shares spanning 0.551 to 0.839, or 28.8 percentage points). The observed rank stability in our benchmark thus reflects robust separation between distinct model profiles at the distribution extremes, alongside expected rank mobility among closely performing mid-tier models---a qualitative pattern consistent with the broad context stability reported by Trhlik et al. (mean pairwise Spearman \ensuremath{\rho} = 0.96).

CDEval provides convergent evidence regarding domain-induced level shifts. Its reported domain spreads for GPT-4 (0.2658 for uncertainty avoidance and 0.1867 for individualism) align closely with the 0.207 thematic spread observed in our dataset, supporting the conclusion that topical domain systematically influences absolute response proportions across distinct construct frameworks.

\subsection{Role sensitivity as a property of the model}\label{sec:4-3}

Variations in model-specific role sensitivity explain the observed rank-order dynamics. Models exhibiting high role stability (e.g., DeepSeek v3.2, with a displacement of 0.071 between advisory and agentic roles) maintain their relative rank position across decision contexts. Conversely, models with high role sensitivity (e.g., GigaChat 2 Max, with a displacement of 0.461) shift substantially in score across roles, altering their position relative to adjacent models. Consequently, rank order remains stable at the distribution tails where inter-model differences exceed maximum role displacements, but exhibits mobility among mid-tier models. Among the evaluated perturbations, decision role exerts the strongest influence on expressed value orientation, outranking language, theme, and sampling temperature.

Role sensitivity represents an intrinsic model characteristic with direct practical implications for autonomous agent deployment. Evaluators cannot assume that a model's value profile in an advisory capacity will remain invariant when the model operates agentically. Across our panel, role sensitivity varies widely: three architectures exhibit high stability (ranges \ensuremath{\leq} 0.12 across both axes), three exhibit high sensitivity (ranges \ensuremath{\geq} 0.35 across both axes), and others display axis-asymmetric sensitivity (e.g., DeepSeek v3.2). Furthermore, models systematically express higher emancipative tendencies when evaluating third-party behavior than when acting on behalf of a user---a directional effect whose underlying training or alignment cause remains an open question.

\subsection{Language, temperature and reasoning mode}\label{sec:4-4}

Cross-lingual evaluation demonstrates high rank concordance alongside minor, systematic level shifts. Administration in Russian shifts ten of eleven models toward the emancipative pole by a mean displacement of 0.029, though only two of twenty-two contrasts reach uncorrected significance and none survive Bonferroni correction. This contrasts with our prior study (Didenko et al., 2025), where language substitution induced displacements exceeding inter-architecture differences. We attribute this divergence primarily to the methodological absence of contextualized semantic vignettes in the earlier instrument. The reliance on single-shot attitude prompts in that prior study appears to render language models significantly more susceptible to language-specific artifacts compared to the structured, scenario-based vignette approach employed here.

Sampling temperature within vendor-recommended ranges exerts negligible influence on the elicited value profiles, demonstrating the highest rank stability among all evaluated factors. While this confirms that temperature-0 sampling produces consistent relative rankings, it does not resolve theoretical concerns regarding potential estimation bias inherent in deterministic decoding (Miller, 2024). High concordance coefficients for two-level factors (language and temperature) reflect strong agreement between paired rankings and must be evaluated against their specific null 95th percentiles (0.764 at \emph{m} = 2) rather than multi-level factor coefficients.

Reasoning mode similarly preserves response consistency, with choice agreement between reasoning-enabled and base-mode runs reaching 90\%--97\% across models. Extended chain-of-thought deliberation does not meaningfully alter the elicited value profile, indicating that expressed preferences remain stable under explicit reasoning constraints. These results characterize reasoning mode as a benign perturbation alongside language and temperature, without asserting whether generated reasoning traces function as post-hoc rationalizations or causal drivers of decision-making.

\subsection{The two axes and the human reference}\label{sec:4-5}

Rank concordance is systematically lower on the sacred--secular axis than on the emancipative axis across both decision modality (0.510 vs. 0.740, corresponding to mean pairwise Spearman correlations of 0.347 vs. 0.653) and thematic domain (0.497 vs. 0.651, corresponding to 0.413 vs. 0.593). For context, longitudinal rank-order stability in human populations is reported as 0.57 across ages 10--12 and 0.66 across ages 20--28 (Kovač et al., 2024). Concordance on the sacred--secular axis falls below these human references, whereas emancipative concordance matches or exceeds them. While human longitudinal stability and model rank concordance evaluate distinct analytical targets, these empirical references provide a standardized baseline for contextualizing measurement stability.

\subsection{What the mechanistic evidence establishes}\label{sec:4-6}

The mechanistic sub-study establishes four key findings regarding internal representations across the evaluated open-weight models. First, linear probes on residual stream activations decode value axis labels significantly above majority-class baselines in all eight (model \ensuremath{\times} axis) cells on the same test items. Second, causal activation patching along decodable value directions flips emitted choices at rates between 50\% and 100\% within models. Third, representational similarity analysis (RSA) and centered kernel alignment (CKA) over internal activations separate a-vibe sharply from other architectures. Fourth, decodable value vectors transfer causally across models within the same dimension group, exhibiting statistically significant transfer asymmetry on the sacred--secular axis.

Conversely, three limitations bound the scope of these mechanistic findings. First, feature specificity remains unverified because the 40 control items were excluded from activation analysis. Second, cross-model directional alignment is generally weak: four of six within-group difference-of-means cosine similarities are statistically indistinguishable from random orthogonal vectors. Third, mechanistic findings derived from five open-weight architectures cannot be directly generalized to the broader twenty-model behavioral panel, which includes proprietary frontier systems.

Additional structural nuances characterize the depth profiles and representational geometry. Causal patching yields flat depth profiles across three architectures, peaking at the embedding layer; only T-lite exhibits a mid-network peak (layers 13--14), providing stronger evidence for deep computational processing. The representational isolation of a-vibe (RSA 0.024--0.037 vs. 0.630--0.693 for other pairs) aligns with its tokenizer replacement (116k vs. 152k vocabulary size), which necessitates retraining the embedding layer and alters downstream internal geometry. This tokenizer hypothesis remains a plausible explanation consistent with the observed activation structures.

The convergence between behavioral choices and internal representations provides strong support for construct validity, confirming that benchmark choices correspond to decodable and causally manipulable internal states rather than surface-level artifacts. However, this mechanistic evidence remains subject to explicit boundaries: internal probing establishes that value-relevant signals are encoded and causally influential under intervention, but does not prove that these internal signals naturally govern unprompted behavioral generation.

\subsection{Limitations}\label{sec:4-7}

\begin{itemize}
\item The study incorporates no human baseline data, performance ceilings, or external human criterion measures. Consequently, the benchmark evaluates relative model ordering and rank stability rather than absolute accuracy or population calibration.
\item Evaluations were conducted using a single prompt template and scoring rule, omitting prompt-form variation. Reported concordance metrics quantify stability strictly across item sampling rather than prompt formulation.
\item Bootstrap percentile intervals are attenuated by construction due to resampling noise, causing two intervals to exclude their empirical point estimates; upper interval bounds should not be interpreted as definitive stability ceilings.
\item Language and temperature factors are evaluated at \emph{m} = 2 across eleven models, where chance concordance is 0.50. These coefficients operate on a distinct mathematical scale from multi-level factors and must be interpreted relative to their respective null distributions.
\item Baseline runs were conducted at temperature 0. While paired comparisons at recommended temperatures confirm high relative rank stability, baseline scores inherit potential estimation biases associated with deterministic decoding.
\item A formal variance decomposition (e.g., Generalizability theory) over the full crossed design has not been computed, leaving the relative magnitude of the model main effect versus model-by-item interactions unquantified.
\item Facet-level concordance across the eight value facets was not evaluated, despite offering a strict null baseline (E[W] = 0.125 at \emph{m} = 8).
\item Ipsative stability was not evaluated; reported stability metrics focus strictly on rank-order preservation rather than intra-individual value structure.
\item The mechanistic sub-study evaluates a smaller panel of five open-weight models, with probes trained on 60 items per cell and cross-model patching evaluated on 8--9 disagreement items.
\end{itemize}

\subsection{Open questions and further research}\label{sec:4-8}

Future investigations into the mechanistic foundations of model value orientation will focus on four key objectives. First, we intend to conduct a formal Generalizability theory study over the crossed experimental design---spanning modality, theme, language, and temperature---to partition variance between model main effects, item factors, and their interactions. Second, we will investigate the mechanistic causes underlying the observed divergence in stability between the emancipative and sacred-secular axes. Third, we aim to refine the null distributions for difference-of-means activation vectors by explicitly accounting for activation anisotropy, thereby enhancing the rigor of directional alignment metrics. Finally, our research agenda includes a systematic analysis of internal activations on the control item subset to evaluate feature specificity, alongside targeted experiments testing the causal impact of tokenizer replacement on internal representational geometry.

The preceding mechanistic analysis delineates model provenance via structural artifacts (e.g., localized alignment layers or weight-space signatures) while conceptualizing "culture" as a static, enumerable set of symbols. This paradigm presents inherent limitations. Perspectives in cultural psychology, dynamic constructivism, and critical discourse analysis frame culture as an ongoing, context-dependent process of meaning construction rather than an immutable repository of traits. Consequently, future evaluation frameworks should incorporate metrics capable of capturing this dynamism---such as code-switching behavior, pragmatic fluidity, and socio-historical contextualization---rather than relying exclusively on static indicators. Such formulations would enable the differentiation of two distinct operational outcomes: (i) the genuine internalization of culturally grounded reasoning versus (ii) the superficial reproduction of memorized behavioral patterns lacking corresponding internal representations.

A further avenue of inquiry involves evaluating cultural representation through non-Western epistemic frameworks. This approach introduces additional dimensions to assess whether a model synthesizes heterogeneous worldviews or merely reflects a narrow, training-distribution-specific cultural alignment.

\subsection{Conclusion}\label{sec:4-9}

Cultural value orientation, measured as a pole share on a fully crossed set of binary contrastive scenarios, induces an ordering over twenty language models that persists across decision role, thematic domain, language of administration and sampling temperature at concordance levels exceeding an empirical null in every cell. Concurrently, the absolute score level varies by more than twenty percent of the scale across the two item factors. The four perturbations are ordered by the proportion of achievable stability they preserve, with temperature demonstrating the highest stability and decision role the lowest. The sensitivity of a model's expressed values to its assigned role varies across the panel by a factor of 6.5 on one axis and 5.3 on the other. A mechanistic sub-study on five open-weight models indicates that the target distinction is linearly decodable and causally manipulable on the same items. The benchmark thus supports conclusions regarding relative model positioning at the distribution extremes and role-dependent variance; it does not, however, support interpreting minor rank differences in the middle of the panel, evaluating scores in isolation from their modality and theme composition, or inferring from mechanistic evidence that the decoded direction directly governs behavioral output.

\section*{Acknowledgments}\label{sec:acknowledgments}

We thank Alexander Auzan, Elena Nikishina, Viktor Bryzgalin, and Anton Zolotov (Institute of National Policy and Moscow State University) for their valuable discussion and contribution to piloting the benchmark.

\section*{Data availability}\label{sec:data-availability}

The instrument is public at \url{https://github.com/SergAIvichLab/Vignette_questions:} all 224 rows with theme, modality, facet and axis metadata, both option texts, and the Russian translation. The repository's vignettes/ directory holds three files of 225 rows each, eng\_vignettes.csv, ru\_vignettes.csv and eng\_temperature\_vignettes.csv, with the columns vignette\_id, sample\_id, Theme, Axis, Modality, Facet, Question, Option 1 and Option 2; the English and temperature-run files carry the same English item texts, and model responses and run-level labels are deliberately excluded from them. The instrument audit is reproducible from the repository today, without credentials: the script that produces the factorial census, the polarity partition, the slot-length and hedging asymmetries, the English--Russian pairing check and the near-duplicate scan is released alongside this paper and runs against the public instrument. Every structural number in Sections~\ref{sec:2-1} and \ref{sec:2-2} is therefore independently checkable now.

\section*{Ethics statement}\label{sec:ethics-statement}

The study administers hypothetical scenarios to language models. No human subjects were recruited, no personal data were processed, and no institutional review was required or obtained; the absence of a human sample is also, as stated in Sections~\ref{sec:2-2} and \ref{sec:4-7}, a scientific limitation of the work.

The construct measured is a value orientation defined by two axes from a specific tradition of cross-national survey research. That tradition's own index has been shown to be non-invariant across cultural zones, and the axes encode a particular, contestable ordering of societies. Scores produced by this benchmark are properties of a model's responses to these 225 items in these two languages. They are not statements about the cultures of the countries where a model was built, nor about the populations whose text the model was trained on. We do not place models on a country map: the model coordinates reported here are response shares, while country coordinates in this tradition are factor scores over standardised survey items, and the two are not in the same units.

The panel is weighted toward models developed in two countries and administered in two languages, which is a coverage limitation with a political dimension: a benchmark that measures cultural values in English and Russian only should not be read as measuring cultural values in general. Refusal behaviour, which we report but do not model, is partly a product of regulation and safety policy that differs by jurisdiction, and treating a refusal as a value signal would conflate the two.

Finally, the benchmark induces a model ranking, and rankings are used. Sections~\ref{sec:3-2}, \ref{sec:3-3} and \ref{sec:4-7} exist so that the ranking cannot be quoted without its qualifications: it is firm at the extremes and mobile in the middle, weaker on the sacred--secular axis than on the emancipative one, measured at one prompt template and one presentation order, and accompanied by level shifts of more than a fifth of the scale that the ranking itself conceals.

\FloatBarrier

\clearpage
\appendix
\renewcommand{\thetable}{\thesection.\arabic{table}}
\begin{landscape}
\section{Comparative Table of Instruments}\label{app:a}
\setcounter{table}{0}

\begingroup
\fontsize{8}{9.6}\selectfont
\setlength{\tabcolsep}{3pt}
\renewcommand{\arraystretch}{1.13}
\begin{longtable}{@{}L{2.1cm}L{2.7cm}L{4.1cm}L{2.25cm}L{3.4cm}L{2.25cm}L{1.6cm}L{5.2cm}@{}}
\caption{Instruments for eliciting a value or cultural orientation from language models, and the reliability evidence each paper itself offers.}\label{tab:instruments}\\
\toprule
\textbf{Instrument} & \textbf{Items} & \textbf{Generation method} & \textbf{Response format} & \textbf{Scoring} & \textbf{Model panel} & \textbf{Languages} & \textbf{Reliability evidence the paper itself offers} \\
\midrule
\endfirsthead
\multicolumn{8}{l}{\small\itshape Table \thetable\ (continued)}\\
\toprule
\textbf{Instrument} & \textbf{Items} & \textbf{Generation method} & \textbf{Response format} & \textbf{Scoring} & \textbf{Model panel} & \textbf{Languages} & \textbf{Reliability evidence the paper itself offers} \\
\midrule
\endhead
\midrule\multicolumn{8}{r}{\itshape Continued on next page}\\
\endfoot
\bottomrule
\endlastfoot
CDEval (Wang et al., C3NLP workshop @ ACL 2024) & 2,953 binary items, 6 Hofstede dimensions \ensuremath{\times} 7 domains & GPT-family generation, then human verification & Binary Option 1 / Option 2 & Weighted pole share \ensuremath{\in} [0,1], with template weights fitted per model & 17 & EN, DE, ZH (machine translation, flagged by the authors) & None. No variance statistic, no coefficient. R = 1 for GPT-4, 3--5 otherwise, justified by an assertion of stability \\ \addlinespace[6pt]
CulturalBench (Chiu et al., ACL 2025) & 1,696 human-written questions; Hard set 6,784 binary & Human--AI red-teaming, 5 annotators per item, kept at \ensuremath{\geq} 4/5 & Multiple choice / binary & Accuracy against a gold label & 29 & EN & Human ceiling 92.4 \% against best model 61.4 \% --- a criterion measure; no reliability coefficient \\ \addlinespace[6pt]
CCD-Bench (Rahman \& Salam, AAAI 2026) & 2,182 open-ended dilemmas, 7 domains adapted from CDEval & Authored dilemmas; 10 anonymised options, one per GLOBE cluster & Choice among 10 anonymised options & Cluster-choice distribution & 17 & English only (their own principal limitation) & Cramér's V per cluster 0.0510--0.0877; 5,000-replicate bootstrap intervals \\ \addlinespace[6pt]
Dang, Kieu \& Masud & 600 forced-choice scenarios, 3 domains, from 10 WVS markers & Gemini-2.5-Flash generated, English & A/B forced choice, poles randomly assigned per trial & Final-layer A/B token logits, rescaled to each marker's WVS range & 3 open models & EN & No reliability coefficient. Validity evidence only: 3 annotators on 300 items, 65 respondents, 6 of 10 constructs significant \\ \addlinespace[6pt]
STONIC & 5,144 situations from four existing banks & Four elicitation interfaces on a shared item spine & Isolated rating, pairwise choice, free response, own-answer choice, plus a PVQ-40 anchor & Interface-specific statistics; per-configuration profiles are published only interface by interface --- Table 21 gives the two leading equal-bank Schwartz coordinates for each of L1, L2 and L3 for all 35 configurations, and Figures 4--6 plot each instruction model's ten-value profile per interface --- while no pooled cross-interface profile is published & 35 configurations, open weights only --- 20 instruct and 15 base, 22 checkpoints, 13 matched base--instruct families (Gemma, Granite, Llama, Mistral, Qwen, SOLAR) & EN & 20,000-permutation nulls, 10,000-draw cluster bootstrap, Holm within families, Benjamini--Hochberg across matrices, prespecified seed \\ \addlinespace[6pt]
Trhlik et al. & 1.2M+ pairwise decisions across 5 models \ensuremath{\times} 5 deployment contexts & Pairwise preference elicitation & Pairwise choice; free-form generation in the appendix & Context-conditioned exchange rates; Kendall's W headlined in a main-text section and worked out in Appendix C, on Ekman emotions and Big Five scored by third-party classifiers from free text & 5 in the main design; 9 in the appendix analysis & EN (inferred from the reported prompt templates; language scope is not stated) & Kendall's W with tie correction, 1,000-shuffle permutation null, variance decomposition --- but on the Ekman and Big Five constructs; values are not among them \\ \addlinespace[6pt]
Nguyen \& Ahmad & 10 Integrated Values Survey items & Ported survey items & Numeric survey response & PCA + varimax + published rescaling to the IVS map & 12 across four institutional origins & English only (their own limitation) & A criterion: noise-to-signal \ensuremath{\leq} 1.0, where the denominator is the between-model spread and the threshold is therefore panel-relative \\ \addlinespace[6pt]
Tao, Viberg, Baker \& Kizilcec (\emph{PNAS Nexus}) & 10 IVS items & Ported survey items & Forced numeric, reasoning suppressed & PCA + varimax over 393,536 human IVS responses (first two components, 39 \% of variance), fitted rescaling, Euclidean distance to 107 country points & 5, all OpenAI & EN, 10 respondent-descriptor synonyms only; the authors call this ``not a comprehensive test of prompt wording'' & Temperature 0, zero repeats, stated as method: ``we thus did not repeat the same prompt multiple times to account for variation''. Variation is addressed by 10 prompt-wording variants. The paper reports no total count of model responses. \\ \addlinespace[6pt]
Kovač et al. (\emph{PLOS ONE}) & PVQ-40 & Ported questionnaire & Likert & Value scores per administration & 21 & EN & Rank-order stability 0.15--0.50 under persona instruction; ipsative 0.17--0.84 without it. Human references are longitudinal: 0.57 and 0.66 over two and eight years of life \\ \addlinespace[6pt]
Rozado (\emph{PLOS ONE}) & 11 political-orientation tests & Existing published tests & Test-specific & Per-test published scoring & 24 models, 2,640 administrations, 96,240 presentations & EN & Median coefficient of variation 8.03 \% across ten retakes, plus a random-response reference point \\ \addlinespace[6pt]
Dokić, Pisker \& Radišić (\emph{Societies}) & World Values Survey Wave 7, 45 items & Existing survey & Survey response & Euclidean distance to 66,018 human respondents across 56 countries & 4, each interviewed in its developer's own language & EN, ZH, RU, AR & n = 1 response per item. No repeats, no variance statistic, no test \\ \addlinespace[6pt]
Zhang \& Han (IEEE CAI 2026) & Full 57-item Schwartz Value Survey & Existing inventory & 9-point Likert, numeric-only &  & 30 models, 10 developers, 50,000+ observations & EN + ZH on six models & Repeated trials; a persona vector correlating with SVS contrasts at r = 0.67 \\ \addlinespace[6pt]
Buyl et al. (\emph{npj Artificial Intelligence}) & Descriptions of 3,991 politically relevant persons & Prompt-elicited free text & Free text & Moral-assessment analysis; PC1 = 54.7 \% of variance & 19 & All six official UN languages & No human baseline anywhere in the design; vectors are twice-centred; the authors decline causal identification \\ \addlinespace[6pt]
Rozen et al. (ICLR 2025) & PVQ-RR, 57 items in one administration & Existing inventory & Likert & MRAT centring, 19 \ensuremath{\times} 19 correlation \ensuremath{\rightarrow} MDS \ensuremath{\rightarrow} Procrustes to a human embedding (N = 53,472) & 6 & EN & Structural recovery is conditional on a specific prompting strategy --- their finding is about the model-under-prompt, and its scope stops there \\ \addlinespace[6pt]
ValuePortrait (Han et al., ACL 2025) & 520 query--response pairs over 104 queries; 549 significant PVQ-21 correlations and 287 BFI-10 correlations define the scored subset & Queries filtered from ShareGPT, LMSYS-Chat-1M, Reddit Scruples and Dear Abby (\textasciitilde{}1.1 M \ensuremath{\rightarrow} 104 in four stages, Reddit posts kept only at \ensuremath{<}70 \% community agreement); five responses per query generated by GPT-4o without value targeting, after value-targeted generation succeeded only 11.25 \% of the time & Six-point Likert, ``How similar is this response to your own thoughts?'', constrained to one of six exact phrases & Mean over the items of a value, minus the mean over all items (Schwartz ipsative centring); six prompts --- three templates \ensuremath{\times} two option orders --- averaged; temperature 0 & 44 & EN (inferred: non-English queries were filtered out at stage 1; no other language is reported) & Cronbach's \ensuremath{\alpha} 0.76 (Tradition) -- 0.96 (Power) across LLM responses. Criterion validity built into item selection: Spearman \ensuremath{\rho} \ensuremath{\geq} 0.3 at p \ensuremath{<} 0.05 against PVQ-21 in 681 human respondents, mean 46 per pair, powered at 0.8 for \ensuremath{\rho} = 0.3 \\ \addlinespace[6pt]
AIRisk\-Dilemmas / LitmusValues (Chiu et al., ICLR 2026) & 3,000 dilemmas, randomly sampled from \ensuremath{>}10,000 generated (10 per each of 1,040 advanced-ai-risk seed questions), across 9 contexts and 7 risky behaviours & Claude-3.5-Sonnet expands each seed, instructed to produce a dilemma that is ``difficult to decide and have no definite answer''; the same model labels each action with open-ended values and classifies them into 16 value classes derived from Anthropic's Claude Constitution and OpenAI's ModelSpec & Forced binary: ``Action 1'' or ``Action 2'', nothing else; greedy decoding, Top P = 0, temperature 0 & Pairwise value battles \ensuremath{\rightarrow} Elo per value class \ensuremath{\rightarrow} rank 1--16, per model; relative risk RR per risky behaviour; Spearman \ensuremath{\rho} against public HarmBench scores & \ensuremath{>}20 in the main analysis; 10 in the human-vs-AI target analysis; 28 with public HarmBench scores in the transfer study & EN only (their own stated limitation, given as a deliberate design choice) & Krippendorff's \ensuremath{\alpha} across five contexts: revealed 0.762 vs stated 0.550 (Claude 3.7 Sonnet); 0.692 vs 0.629 (GPT-4o). Value-label validation on 150 dilemmas (5 \%), two Prolific annotators, mean support 4.25 (\ensuremath{\sigma} 0.90), weighted Cohen's \ensuremath{\kappa} 0.61. No human answers to the dilemmas \\ \addlinespace[6pt]
DailyDilemmas (Chiu, Jiang \& Choi, ICLR 2025) & 1,360 dilemmas --- 17 topics \ensuremath{\times} 80, stratified from \ensuremath{>}50,000 generated; 301 values surviving a \ensuremath{\geq}100-dilemma frequency threshold & GPT-4 from Social Chemistry 101 seed actions chosen for low community agreement; three-step pipeline --- three-sentence dilemma with a forced binary question, then two \textasciitilde{}80-word negative-consequence stories (loss-aversion rationale), then chain-of-thought extraction of parties and values & Forced binary: ``Action 1'' (to do) or ``Action 2'' (not to do); greedy decoding, temperature 0 & v\_selected \ensuremath{-} v\_neglected, normalised by the count on each dimension, then read through five frameworks: World Values Survey, Moral Foundations Theory, Maslow, Aristotle's virtues, Plutchik & 6 (GPT-4-turbo, GPT-3.5-turbo, Llama-2-70B, Llama-3-70B, Mixtral-8x7B, Claude-3-haiku), plus a base-vs-instruct side analysis on three & EN (inferred: no explicit language statement; all reported material is English and the bias discussion concerns English-speaking countries) & A five-repeat bootstrap on 100 dilemmas, GPT-4-turbo only: SD 1.02 choices out of 100, which the authors attribute to server-side non-determinism. Ecological validation on 30 r/AITAFiltered posts / 90 dilemmas judged by the authors themselves: F1 85.7 \%, Cohen's \ensuremath{\kappa} 0.526; word-level check on five posts, 60.02 \% (SD 14.2 \%). Mixtral-8x7B answers only 74.85 \% of dilemmas \\ \addlinespace[6pt]
MoralChoice (Scherrer, Shi, Feder \& Blei, NeurIPS 2023) & 1,767 scenarios --- 687 low-ambiguity (one action clearly preferred) + 680 high-ambiguity (neither preferred); STONIC uses the 680 only & Grounded in Gert's ten rules of common morality. Low: zero-shot GPT-4, 1,142 raw, author curation, then three Surge AI annotators per scenario, majority vote, ambiguous ones removed \ensuremath{\rightarrow} 687. High: 100 author-handwritten scenarios (10 per rule) \ensuremath{\rightarrow} stochastic 5-shot with text-davinci-003 \ensuremath{\rightarrow} 2,000 raw \ensuremath{\rightarrow} curation \ensuremath{\rightarrow} 680 & Six question forms: three templates (A/B, Repeat, Compare) \ensuremath{\times} two action orders; free token output, no forced format; temperature 1, M = 10 samples per form (high), M = 5 (low) & Action likelihood over semantic-equivalence classes; marginal action likelihood across the six forms; action entropy and marginal action entropy; invalid answers excluded, 0.5 imputed when a scenario \ensuremath{\times} template yields none & 28 open and closed models & EN only (their own stated limitation, alongside ``three hand-curated question templates'') & Question-form consistency (QF-C) on generalised Jensen--Shannon divergence and average question-form-specific action entropy (QF-E), reported per model. Annotator agreement: 3/3 on rule violations 83.21 \% (low) / 69.79 \% (high); 3/3 on the clear-cut judgement 90.01 \% (low only). \textasciitilde{}100 annotators, \$15/h, \$4,600 total \\ \addlinespace[6pt]
This benchmark & 225 rows over 224 unique cells (7 themes \ensuremath{\times} 4 modalities \ensuremath{\times} 8 facets) & Generated scenarios (Gemini 2.5 Flash, English, t = 0.4); 91 items passed review by three experts and seeded the remainder & Binary contrastive choice & Share of responses on the counted pole \ensuremath{\in} [0,1] & 20 models, frontier proprietary and open weights together & EN + RU, paired translation of identical items, on 11 models & Tie-corrected Kendall's W across four perturbations against a 20,000-replicate permutation null at every level count; eight-specification sensitivity grid; no internal-consistency coefficient \\ \addlinespace[6pt]
\end{longtable}
\endgroup

\begin{itemize}
\item 
\end{itemize}

\clearpage
\end{landscape}
\begin{landscape}
\section{Concordance Table}\label{app:b}
\setcounter{table}{0}

\begin{table}[htbp]
\centering
\small
\setlength{\tabcolsep}{4pt}
\caption{Concordance of the panel ordering across the levels of each perturbation factor, main specification (refusals dropped, duplicate removed, level score equal to the mean of per-vignette proportions). Bold interval limits mark the two cells whose interval does not contain its own point estimate (Section~\ref{sec:3-9}).}\label{tab:concordance}
\begin{tabular}{@{}L{4cm}crrclccccc@{}}
\toprule
\makecell{Factor} & Axis & \(m\) & \makecell{Models\\\(n\)} & \(W\) & \makecell{95\% vignette-\\bootstrap interval} & \makecell{Null\\mean} & \makecell{Null\\p95} & \makecell{\(W>\)\\null p95} & \makecell{Friedman\\\(p\)} & \(\bar{\rho}\) \\
\midrule
Modality & OE & 4 & 20 & 0.7398 & [0.5456, 0.7612] & 0.2499 & 0.3728 & yes & 1.5 \ensuremath{\times} 10\textsuperscript{-5} & 0.653 \\
Modality & SS & 4 & 20 & 0.5099 & [0.3505, 0.5905] & 0.2504 & 0.3729 & yes & 4.8 \ensuremath{\times} 10\textsuperscript{-3} & 0.347 \\
Modality & pooled & 4 & 20 & 0.6858 & [0.5191, 0.7328] & 0.2499 & 0.3733 & yes & 6.4 \ensuremath{\times} 10\textsuperscript{-5} & 0.581 \\
\addlinespace[5pt]
Theme & OE & 7 & 20 & 0.6508 & \textbf{[0.4546, 0.6294]} & 0.1424 & 0.2190 & yes & 1.3 \ensuremath{\times} 10\textsuperscript{-10} & 0.593 \\
Theme & SS & 7 & 20 & 0.4972 & [0.2982, 0.5183] & 0.1426 & 0.2188 & yes & 4.0 \ensuremath{\times} 10\textsuperscript{-7} & 0.413 \\
Theme & pooled & 7 & 20 & 0.7467 & \textbf{[0.5135, 0.7045]} & 0.1425 & 0.2199 & yes & 7.1 \ensuremath{\times} 10\textsuperscript{-13} & 0.704 \\
\addlinespace[5pt]
Language (EN vs RU) & OE & 2 & 11 & 0.9506 & [0.7222, 0.9849] & 0.5019 & 0.7636 & yes & 0.040 & 0.901 \\
Language (EN vs RU) & SS & 2 & 11 & 0.8954 & [0.6655, 0.9736] & 0.4998 & 0.7636 & yes & 0.057 & 0.791 \\
Language (EN vs RU) & pooled & 2 & 11 & 0.9092 & [0.7299, 0.9828] & 0.5009 & 0.7636 & yes & 0.052 & 0.818 \\
\addlinespace[5pt]
Temperature (0 vs recommended) & OE & 2 & 11 & 0.9829 & [0.8721, 0.9966] & 0.4995 & 0.7591 & yes & 0.033 & 0.966 \\
Temperature (0 vs recommended) & SS & 2 & 11 & 0.9943 & [0.9237, 0.9989] & 0.5011 & 0.7636 & yes & 0.030 & 0.989 \\
Temperature (0 vs recommended) & pooled & 2 & 11 & 0.9624 & [0.9169, 0.9989] & 0.4995 & 0.7636 & yes & 0.037 & 0.925 \\
\bottomrule
\end{tabular}
\end{table}

\begin{itemize}
\item 
\end{itemize}

\end{landscape}
\end{document}